\documentclass[10pt,letterpaper,twocolumn]{article}

\usepackage{usenix-2020-09}
\usepackage{silence}
\usepackage{amsmath,amsopn,amssymb}
\usepackage{subfig}
\usepackage{endnotes,microtype,xspace,graphicx,fancyvrb,multirow}
\usepackage{booktabs}
\usepackage{array,underscore,relsize}
\usepackage[T1]{fontenc}
\usepackage{times}
\usepackage{fancyhdr,lastpage}
\usepackage{enumitem}
\usepackage[labelfont=bf,font=small,skip=5pt]{caption}
\usepackage[compact]{titlesec}
\usepackage{titling}
\posttitle{\par\end{center}\vskip 0.5em}
\posttitle{\par\end{center}}
\usepackage{fp}
\usepackage{siunitx}

\usepackage[draft]{minted}

\usepackage{balance}

\usepackage{comment}

\usepackage{pifont}% http://ctan.org/pkg/pifont

\usepackage[vlined,linesnumbered]{algorithm2e}
\usepackage{amsmath}

\usepackage{bookmark}

\newcommand{\round}[2]{\num[round-mode=places,round-precision=#1]{#2}}

\newcommand{\sys}{\mbox{\textsc{PAL}}\xspace}

\newcommand{\XXX}[1]{\textcolor{red}{XXX: #1}}

\newcommand{\cc}[1]{\mbox{\smaller[0.5]\texttt{#1}}}

\fvset{fontsize=\scriptsize,xleftmargin=8pt,numbers=left,numbersep=5pt}

\input{code/fmt}
\newcommand{\figrule}{\hrule width \hsize height .33pt}
\newcommand{\coderule}{\vspace{-1.4em}\figrule\vspace{0em}}

\def\Snospace~{\S{}}

\if 0

\fi

\newif\ifdraft\drafttrue
\newif\ifnotes\notestrue
\ifdraft\else\notesfalse\fi

\input{glyphtounicode}
\newcolumntype{R}[1]{>{\raggedleft\let\newline\\\arraybackslash\hspace{0pt}}p{#1}}

\newcommand{\squishlist}{
\begin{itemize}[noitemsep,nolistsep]
  \setlength{\itemsep}{-0pt}
}
\newcommand{\squishend}{
  \end{itemize}
}

\usepackage{tikz}
\newcommand*\WC[1]{%
\begin{tikzpicture}[baseline=(C.base)]
\node[draw,circle,inner sep=0.2pt](C) {#1};
\end{tikzpicture}}

\newcommand*\BC[1]{%
\begin{tikzpicture}[baseline=(C.base)]
\node[draw,circle,fill=black,inner sep=0.2pt](C) {\textcolor{white}{#1}};
\end{tikzpicture}}

\usepackage{xstring}
\newcommand{\PP}[1]{
\vspace{2px}
\noindent{\bf \IfEndWith{#1}{.}{#1}{#1.}}
}

\newcommand{\boxbeg}{
\vspace{2px}
\noindent\begin{tabular}{|l|}\hline
\begin{minipage}{3.2in}
\vspace{2px}
\noindent
}

\newcommand{\boxend}{
\vspace{2px}
\end{minipage}\\ \hline
\end{tabular}
\vspace{-3pt}
}

\newcommand{\boxrule}{
  \vspace{-7pt} \par\noindent\rule{\textwidth}{0.4pt}
}  

\gdef\therev{5191ff8}
\gdef\thedate{2021-12-14 11:30:32 +0900}

\begin{document}

\title{\vspace{-30pt}\textbf{\LARGE In-Kernel Control-Flow Integrity on Commodity OSes \\
using ARM Pointer Authentication}}

% when 'make draft'
\ifdefined\DRAFT
 \pagestyle{fancyplain}
 \lhead{Rev.~\therev}
 \rhead{\thedate}
 \cfoot{\thepage\ of \pageref{LastPage}}
\fi

%\author{Paper \#287\vspace{-3em}}

\author{
 	Sungbae Yoo$^\dagger$$^\ddagger$\;
 	Jinbum Park$^\dagger$$^\ddagger$\;
 	Seolheui Kim$^\dagger$\;
 	Yeji Kim$^\dagger$\;
 	Taesoo Kim$^\dagger$$^\ast$\;
 	\\\\
 	\emph{$^\dagger$ Samsung Research}, \\
 	\emph{$^\ast$ Georgia Institute of Technology}
}

%% \author{
%%  Isaac Newton$^\dagger$\;
%%  Albert Einstein$^\ddagger$\;
%%  Marie Curie$^\ast$\;
%%  James Clerk Maxwell$^\dagger$\;
%% \\
%%  Richard Feynman$^\dagger$\;
%% \\\\
%%  \emph{$^\dagger$  Trinity College}, \\
%%  \emph{$^\ddagger$ University of Zurich},
%%  \emph{$^\ast$       University of Paris}
%% }

\date{}
\maketitle

\sloppy

\begin{abstract}
  This paper presents
  an in-kernel, hardware-based control-flow integrity (CFI) protection,
  called \sys,
  that utilizes ARM's Pointer Authentication (PA).
  It provides three important benefits
  over commercial, state-of-the-art PA-based CFIs
  like iOS's:
  1) enhancing CFI precision via
     automated refinement techniques,
  2) addressing hindsight problems of PA for in-kernel uses
  such as preemptive hijacking and brute-forcing attacks,
  and 3) assuring the algorithmic or implementation correctness
  via post validation.

  \sys achieves these goals in an OS-agnostic manner,
  so could be applied to commodity OSes like
  Linux and FreeBSD.
  The precision of the CFI protection
  can be adjusted for better performance
  or improved for better security
  with minimal engineering efforts
  if a user opts in to.
  Our evaluation shows that
  \sys incurs negligible performance overhead:
  e.g., <1\% overhead for Apache benchmark
  and 3--5\% overhead for Linux perf benchmark
  on the latest Mac mini (M1).
  Our post-validation approach
  helps us ensure the security invariant
  required for the safe uses of PA inside the kernel,
  which also reveals new attack vectors
  on the iOS kernel.
  %
  % We plan to commercially deploy this implementation
  % on the consumer electronic devices
  % (e.g., home appliances and smart phones).
  %
  \sys as well as the CFI-protected kernels
  will be open sourced.
\end{abstract}

\section{Introduction}
\label{s:intro}

\let\thefootnote\relax\footnote{$^\ddagger$ The authors contributed equally.}

%
% An important observation is that the vtable pointers themselves have no address
% diversity; they're signed with a zero-context. This means that if we can
% disclose a signed vtable pointer for an object of type A at address X, we can
% overwrite the vtable pointer for another object of type A at a different address
% Y.
%
% This might seem completely trivial and uninteresting but remember: we only have
% a linear heap buffer overflow. If the vtable pointer had address diversity then
% for us to be able to safely corrupt fields after the vtable in an adjacent
% object we'd have to first disclose the exact vtable pointer following the object
% which we can overflow out of. Instead we can disclose any vtable pointer for
% this type and it will be valid.
%
% The clang design doc explains why this is:
%     It is also known that some code in practice copies objects containing v-tables
%     with memcpy, and while this is not permitted formally, it is something that may
%     be invasive to eliminate.
%
% Right at the end of this document they also say "attackers can be devious." On
% A12 and above we can no longer trivially point the vtable pointer to a fake
% vtable and gain arbitrary PC control fairly easily. Guess we'll have to get
% devious :)

Memory safety issues are the foremost security problems
in today's operating systems---%
in 2020 alone, there were 149 CVEs assigned
to potentially exploitable bugs in Linux~\cite{cve-linux}.
To prevent latent bugs from exploitation,
commodity OS vendors
have been developing and deploying
modern mitigation techniques
such as KASLR, DEP and SMA/EP (PXN).
However,
more powerful exploitation techniques,
such as return- and jump-oriented programming~\cite{roemer2012return, jop},
have been developed and demonstrated that
such migration schemes can be
ultimately bypassed~\cite{pwn2own}.
As a response,
control-flow integrity (CFI)~\cite{cfi, cfi-survey},
which enforces a program's control transition
(e.g., an indirect call or a return)
to strictly follow the known control graph
curated at compilation time,
has been considered
as a promising, necessary direction
to mitigate these emerging exploitation techniques.
Accordingly,
modern operating systems like Android, Windows and iOS
all implement
some forms of CFI~\cite{android-cfi,androidblog-cfi,ms-cfg,apple-cfi}.

During the last several years,
there has been exhaustive research exploration
of CFI's design space~\cite{cfi-survey},
which falls broadly into two categories:
\WC{1} enhancing the precision of CFI
(i.e., reducing the number of targets that an indirect call can take);
and \WC{2} making CFI protection faster and practical
(i.e., incurring minimum CPU and memory overheads).
The community has improved CFI precision
by providing better algorithmic advances
to model control-flow transitions accurately%
~\cite{ctx-sensitive-analysis,flow-sensitive-analysis},
or by utilizing exact run-time contexts~\cite{ding:pittypat,hu:ucfi}.
% one could successfully
% estimate the indirect call targets into one
% at run-time~\cite{hu:ucfi}.
%
However, in practice,
the performance overhead often
determines the feasibility
of actual deployment---%
it would be acceptable to prevent the most common cases
with negligible overhead
rather than fully preventing all of them with obtrusive overhead.
One recent approach
taken by Apple~\cite{apple-cfi} and researchers~\cite{pacitup,patter}
is to speed up CFI
by utilizing hardware-based protection,
called ARM Pointer Authentication (PA).

In this paper,
we propose yet another in-kernel CFI protection based on PA,
called \sys, which aims to enhance CFI precision (see \autoref{table:allowed-target}) in an automated manner
while imposing negligible performance overhead (see \autoref{s:eval:performance}).

More specifically,
we implement a context analyzer (see \autoref{ss:context-analyzer})
that captures common idioms and design patterns in commodity OSes (see \autoref{ss:context})
to enhance the CFI precision.
OS developers just can run the context analyzer with the desired CFI precision level,
then it automatically produces kernel code with high CFI precision.
For example,
the context analyzer can refine the indirect call targets
based on the invariants of a kernel object
% (e.g., associating a device name field with a member pointer as a part of PA's context)
or based on a calling context of a kernel API.
%And it is OS-agnostic so that it could be easily applied to Linux and FreeBSD.
% (e.g., limiting the invocation of a callback passed to a \cc{kref} call in Linux).
%
% Furthermore, this approach makes \sys still be compatible to memory copy functions
% without sacrificing security
% (see~\autoref{t:context-change}).
% In contrast,
% C++/Object-C V-Table pointers
% on iOS
% make it fundamentally difficult
% to design in a secure manner~\cite{ios-mac-security}
% as it is recently demonstrated vulnerable~\cite{google-zero-context}.
% %
% In addition, we provide a tool to help developers
% for correct annotations, which could prevent human errors at the first stage
% (see \autoref{ss:context-analyzer}).

On top of that, unlike other PA-based schemes
that place the compiler and their underlying algorithms as TCB,
we introduce a static validator(see \autoref{ss:validator})
that can recognize common pitfalls
and attack vectors of PA
on the \emph{final} kernel image.
Such separation of concerns helps us
develop PA-based CFI and refinement rules
in a higher-level, early-stage IR (i.e., GIMPLE in the GCC)
without dealing with the subtleties
of the back-end compiler optimizations.
In other words,
as long as the static validator assures that
a certain image has no pitfall,
we know that
the final kernel image meets all the invariants
necessary for the secure enforcement of \sys's CFI at run-time.
For example,
we commonly see that
a later-stage optimization
makes
a PA-protected pointer from an early stage
out of a for-loop and stashes into a memory,
disarming the PA-based mitigation
(see~\autoref{s:threat} and Google's PoC on iOS~\cite{google-project-zero}).

\sys's PA-based approach
exhibits better CFI precision
than any other commercial solution:
compared with Android's CFI~\cite{androidblog-cfi},
the number of indirect calls having >100 targets reduces
from 7.0\% to 0.08\%.
% and,
% compared with iPhone's PA-based CFI~\cite{google-project-zero},
% \XXX{revise: no using the term context yet. need concise wording.}
% the number of contexts used for more than 100 function pointers
% reduces from 21.2\% to 0\%.
%
Most importantly, to the best of our knowledge,
\sys is the first PA-based scheme that has successfully evaluated
performance on a real PA-supported hardware (the Mac mini based on the M1 chip).
Our evaluation shows
that \sys imposes negligible performance overhead
on user applications:
<1\% in the Apache benchmark,
while incurring 6.8\% kernel-space overhead on average in the LMbench's benchmarks that had measurable overhead.
(i.e., benchmarks with over 1~$\mu$s overhead)

This paper makes the following contributions:

\squishlist
\item \textbf{New attack vectors.}
  We provide a systematic categorization
  of attack vectors
  under a precise yet powerful definition of threat models
  when using ARM PA to protect the OS kernel.
\item \textbf{Automated refinement techniques.}
  We implement a context analyzer
  to capture idioms and design patterns
  that enhance the precision of the CFI protection
  in automated yet OS-agnostic ways.
  We demonstrate our context analyzer on two commodity OSes:
  Linux and FreeBSD kernels.
\item \textbf{Static validator.}
  We implement a static validator
  to automatically verify
  non-trivial security problems
  unintentionally introduced by complex compiler's optimization
  as well as implementation mistakes.
  It found new attack vectors in the latest iOS kernel
  as well as bugs in other PA-based schemes.
\item \textbf{Open source.}
  We will make the end-to-end ecosystem open source
  including compiler plugin, static validator,
  the CFI-protected Linux and FreeBSD kernels, and context analyzer.
\squishend

\section{Background}
\label{s:background}

\begin{figure}
	\includegraphics[width=\columnwidth]{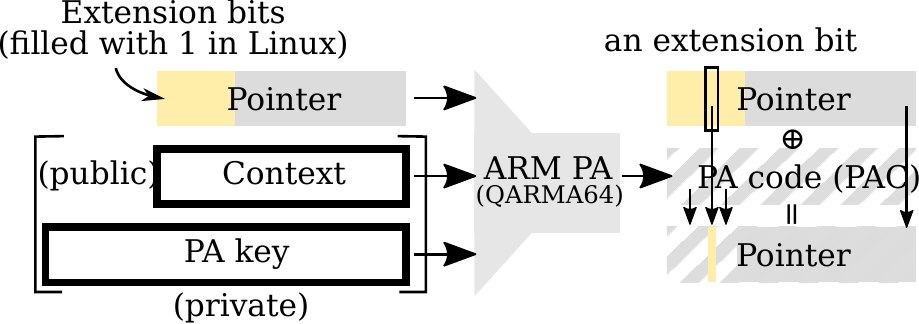}
	\caption{A PA code (PAC)
      is generated
      using QARMA64~\cite{qualcomm-pac} block cipher
      with three inputs: a pointer, a context (public) and a PA key (private).
      And the PA-signed pointer
      contains the PAC
      in the prefix of the pointer
      that indicates its authenticity~\cite{mosec-pac}.
      The signed function pointer will be copied or moved around
      together with the PAC, and checked its authenticity
      right before uses.}
	\label{fig:xpacGadget}
\end{figure}

\subsection{Control-Flow Integrity (CFI)}
As modern defense schemes
like Data Execution Protection (DEP)
prevent \emph{code injection} attacks,
offensive techniques
like return- or jump-oriented programming (ROP/JOP)~\cite{roemer2012return,jop}
have been proposed
to reuse existing code snippets
for exploitation,
commonly known as \emph{code-reuse attacks}.
With ROP and JOP techniques,
an attacker can perform arbitrary computation~\cite{roemer2012return}
by chaining existing code gadgets
\emph{after} hijacking a program's control flow
(e.g., overwriting a function pointer or a return address).
Since the code gadgets end
with a return or a jump instruction
that allows further chaining,
they are called \emph{return-} or \emph{jump-}oriented programming.

One promising mitigation to prevent
such code-reuse attacks
is to ensure that
the integrity of control-flow transfers
remains intact
(i.e., making an intended transition)
at run-time, so called control-flow integrity (CFI).
%
% One simplest example is
% to check
% an integrity of a function call and its return---%
% the transition from a callee to a caller
% can be checked if integral
% to prevent stack smashing attacks~\cite{phrack-stack-smashing}.
%
The most common approach is to
extract control-flow graphs (CFG)
from the source code during compilation
and validate that all transitions
are legitimate at run-time
(i.e., following an edge of the extracted CFG).
As a standard CFG is overly conservative---%
an indirect jump instruction
can lead to any location in a program's memory,
CFI often utilizes high-level structures
from programming languages.
To limit indirect calls,
there are different strategies for identifying
what is the set of functions that can be involved at each call site :
1) each function's entry~\cite{ms-cfg},
2) a group of functions with the same types in \cc{C}~\cite{android-cfi,pax-rap},
3) virtual functions in the class hierarchy in \cc{C++}~\cite{android-cfi,vtable}.
As most transitions depend on the program's execution state,
dynamic approaches that attempt
to associate the execution state and transition
are proposed,
e.g., using inputs~\cite{picfi},
shadow execution/traces~\cite{ding:pittypat},
and collecting control-sensitive information~\cite{hu:ucfi}.
The precision of these dynamic approaches, however,
comes with non-trivial run-time overheads and
hindering deployment in practice.

Common CFI schemes mostly concern
\emph{forward} transitions (i.e., an indirect call)
while relying on \emph{backward} transitions (i.e., a function return)
to be protected by other orthogonal techniques
such as a shadow stack~\cite{intel-cet, shadowstack}.
In \sys,
we aim to design an end-to-end solution
that protects \emph{both} forward and backward edges.

% ARM's PA also provides
% a simple form of backward-edge protection
% by signing the return address~\cite{qualcomm-pac}

\subsection{Pointer Authentication}
\label{s:arm-pac}

ARMv8.3-A introduced
a new hardware-based security feature,
called \emph{Pointer Authentication} (PA),
that can check the integrity of a pointer
with minimal performance
(i.e., using the fast QARMA64 block cipher)
and storage overhead
(i.e., using the extended bits of a pointer).
Simply put,
a pointer is signed (\cc{PAC}) when generated
and is authenticated (\cc{AUT}) before its use,
similar in concept to tagged memory~\cite{mem-tag}.

\vspace{2px}
\noindent \BC{1} \textbf{\cc{PAC}}
\emph{signs} a function pointer
with a signing key and a context as a nonce (see~\autoref{fig:xpacGadget}).
Since the ARM 64-bit processor does not fully utilize
the entire 64-bit address space (using 39-48 bits in Aarch64),
the signed pointer
can carry the authentication code (PAC)
as part of the pointer (25-16 bits in the extension bits).
This design decision simplifies
the heavy bookkeeping required
to propagate the PAC for pointers---now,
a normal instruction like \cc{mov}
seamlessly propagates the PAC
of a pointer in a register
without additional costs.

PA provides five registers for encryption keys
that can only be set in privilege mode:
APIA and APIB for code pointers,
APDA and APDB for data pointers,
and APGA for general-purpose use,
each of which is used
to compute a cryptographic hash (i.e., QARMA64~\cite{qualcomm-pac})
to generate a Message Authentication Code (MAC).

\vspace{2px}
\noindent \BC{2} \textbf{\cc{AUT}}
\emph{authenticates}
the signed pointer
and restores it to its original form
(i.e., discarding the PAC and restoring the extension bits)
given the original key and context.
If the authentication succeeds,
the restored function or data pointer
is used as intended
by the following instructions.
However, when the authentication fails,
\cc{AUT} simply \emph{flips an error bit} in the pointer
to indicate the corrupted state,
and any \emph{later use} of the pointer
raises an exception,
as the restored function pointer
does not conform to the canonical form
of the virtual address space.

\PP{Using context}
PA's context is a critical element
to narrow down
the protection domain
because all function pointers signed by the same context
can be used alternatively in an indirect call
(note, all signed by the same PA key as long as
PA key is not changed).
There are two well-known choices of contexts:
a stack pointer (\cc{sp}) as a context
to sign a return address
(i.e., a backward CFI),
and zero as a context
to sign all function addresses,
which does not need to identify all call-sites and
their functions to be involved.
More advanced uses,
like using a type signature as a context~\cite{pacitup},
also not requiring to identify relations
between all call-sites and functions,
have also recently been proposed,
but our focus in this paper
is to refine the context
by using \emph{dynamic} information
that can be captured from
design idioms in commodity OS kernels.

\PP{EnhancedPAC2 and FPAC}
ARM recently announced
two new features, namely, EnhancedPAC2 and FPAC~\cite{arm-8-6}
to address problems found by Google~\cite{google-project-zero}.
EnhancedPAC2 changes PAC bits to be larger
by XOR-ing with the upper bits of the pointer,
helping PA to avoid brute-forcing attacks.
FPAC makes an \cc{aut} instruction raise
an exception immediately
upon the authentication failure.
With FPAC, \sys can optimize the performance even further,
but our current focus is to be compatible
to ARM v8.3-A,
which will be available on the consumer market soon.
Such optimization techniques
based on these hardware features
can also be used to improve
the performance of \sys.
\section{Overview}
\label{s:overview}

\subsection{Threat Model}
\label{s:threat}

We assume an attacker has the capabilities of
arbitrary reads and writes
\emph{at arbitrary moments},
similar to PaX/RAP~\cite{pax-rap}.
We also assume that
the victim has
all modern defense mechanisms
deployed, namely,
secure boot, stack protection, DEP, KASLR, PXN,
and page-table protections~\cite{tzrkp, skee}
to protect the kernel's non-writable regions
from page-table modification attacks such as KSMA~\cite{ksma}.
It is worth noting that
the capabilities of arbitrary reads and writes
do not mean that it is possible to inject code
so that existing code snippets should be reused for attack.
Lastly, we assume KASLR can be bypassed
either by inferring the layout of the kernel image~\cite{jang:drk-ccs, klasr-side-channel}
or via common information leakage~\cite{xu:deadline}
otherwise
an arbitrary write or read
is prevented in the first place.

In the kernel,
arbitrary read/write capabilities
implicate more than just crafting memory;
since an attacker can force
all registers to be spilled to memory
via preemption,
an attacker can modify all values of the \emph{registers}
stored in memory (i.e., execution context).
This model would be pessimistic,
as real-world vulnerabilities
typically allow limited capabilities
like restricted overflow,
but we believe this is the right way
to reason about the strong security guarantees
of the defense mechanisms
in the kernel~\cite{pax-rap}.

\PP{Control-flow hijacking}
We assume the primary goal
of an attack is to hijack
the control-flow of the kernel
and then, leverage it to launch
post-exploitation payloads
such as obtaining a root shell,
exfiltrating information, installing a rootkit, etc.
With arbitrary memory reads/writes,
this is not the only approach
(e.g., data-oriented attack~\cite{dfi,song:hdfi}),
but still is the most prevalent, reliable, and stealthy form
of powerful attack~\cite{cfi-survey}.

\PP{Out-of-scope attacks}
We do not consider side-channel attacks
such as micro-architectural~\cite{meltdown,spectre},
timing~\cite{armageddon,flush-reload},
or electromagnetic side-channels~\cite{electromagnatic},
because they are mostly limited to secrecy violation (i.e., information leak).
Similarly,
hardware attacks
such as Rowhammer~\cite{rowhammer} caused by faulty DRAM
are not of our concern.
All high-privilege components
like the hypervisor, firmware and hardware
are our TCB.
We also do not consider any advanced forms
of data-oriented attacks~\cite{dfi}
but have a plan to extend our approach
to mitigate prevalent forms of data-oriented attacks,
similar to HDFI~\cite{song:hdfi}, by using PA.
%
%Our concern and defense in this paper
%are orthogonal
%of what these types of attacks and defenses aim to achieve.

\PP{Correctness assumption}
We do not rely on the correctness of complex compilation tool-chains
that have sophisticated optimization algorithms in their back-ends.
Instead, we trust
a static post validator
that directly analyzes the \emph{final} kernel image
and ensures that the invariants required for the secure uses of PA
are correctly respected in the binary
(see~\autoref{ss:validator}).

\subsection{Attack Vectors against ARM PA}
\label{s:av}

We categorize
the fundamental limitations of ARM PA
into two classes,
namely, pointer substitution attacks (\BC{1}, \BC{2}, \BC{3})
and improper uses of the PA protection (\WC{1}, \WC{2}).

\vspace{2px}
\noindent \BC{1} \textbf{Replaying attack}.
\label{s:replaying-attack}
A leaked, signed function pointer
can be legitimately reused
in any indirect call
with the same context.
This means
an adversary having arbitrary read capability
can scan the entire kernel memory
and collect outstanding function pointers
for the replaying attack.
This fundamental problem can be quantified
by measuring
the number of function pointers signed by the same context
and the number of indirect call sites
authenticating pointers
with the same context (\autoref{s:eval:security}).
To reduce the substitution targets,
any PA-based protections should minimize
the uses of the same contexts.

\noindent \BC{2} \textbf{Forging pointers via signing gadgets}.
\label{s:forging-pointers}
Instead of passively scrubbing function pointers,
an adversary can generate
a signed pointer
with the context of his/her own choice
for the targeted indirect call.
There are two prevalent situations:
1) an attacker hijacks
a stored function pointer or the context
during the signing process
and 2) an attacker provides
an arbitrary function pointer
to a re-signing routine
that ignores authentication failures.

\vspace{2px}
\noindent \BC{3} \textbf{Brute-forcing attack}.
\label{s:brute-force-attack}
PA reserves a small number of bits
(e.g., 15 bits in the 48-bit address space)
to embed a MAC as a part of the pointer.
Unfortunately,
this is so small
that an attacker can identify the correct MAC
by enumerating all the input space
if there exists a proper oracle.
This problem is particularly difficult to mitigate
because the production kernel
cannot simply panic
when an authentication failure happens,
whether that is a malicious attempt or not~\cite{linus-security}.
All the existing PA-based protections
suffer from this attack vector.

\vspace{2px}
\noindent \WC{1} \textbf{Key leakage and cross-EL attack}.
\label{s:cross-el-attack}
ARM~\cite{arm-reference} and Qualcomm~\cite{qualcomm-pac}
specify PA's behaviors
only in the context of the user space.
To utilize PA for kernel protection,
according these references,
the PA keys should be multiplexed (or virtualized)
for both user and kernel spaces.
However, under our threat model,
an adversary can sign any function pointers in user space
by using the same key and context as the kernel,
a so called cross-EL attack.
Any preventive measures that keep track of PA keys in the kernel space
should be cautious about
\emph{not storing} the key to memory---%
our adversary with the arbitrary read capability
can obtain the PA key.

\vspace{2px}
\noindent \WC{2} \textbf{Time-of-check to time-of-use (TOCTOU)}.
\label{s:preemption-hijacking}
PA does not guarantee atomicity of its check and use,
meaning that there is a time window
between two PA instructions.
There are two problems that can occur
during this time window:
1) a raw pointer is unintentionally spilled before a use
mainly due to
later-stage compiler optimizations or machine-code generation,
and 2) an attacker enforces a preemption right
before its use,
causing a raw pointer stored in the register
to be spilled to an attacker-controlled memory.

One naive solution is to disable an interrupt
during this time window
but we observed an important drawback:
it increases the interrupt-disabled regions and will finally
turn out unacceptable due to performance fall-off.
(e.g., a virtual call in the VFS layer will disable the interrupt for the rest of executions, which drops concurrency of each file operation).
Any in-kernel PA defenses
should prevent this attack vector
without introducing complexity and performance overheads.

\begin{comment}
Authentication and use are not atomic in PA;
this means
there is a time window between
the function pointer authentication
and its use.
%
Two types of problems can occur during this time window:
1) an authenticated pointer (a raw pointer)
is unintentionally spilled before use,
largely due to later-stage compiler optimizations (see~\autoref{fig:validator-p2}),
and 2) an attacker enforces preemption right
after the authentication, so a raw pointer
stored in the register has to be spilled to memory
where the attacker can read and write.

One naive solution is to disable an interrupt
between the authentication and use,
but we observed two important drawbacks:
1) it increases the interrupt-disabled regions unacceptably
because the interrupt should be disabled during any indirect calls
(e.g., a virtual call in the VFS layer will disable the interrupt for the rest of executions),
and 2) it requires a separate bookkeeping to track its nested state
to support subsequent indirect calls,
similar to a preemption counter.
% \footnote{non-returning functions like signal make this approach even more complicated.}.
%
Any in-kernel PA defenses
should prevent this attack vector
without introducing complexity and performance overheads.
\end{comment}

\section{Design}
\label{s:design}

\begin{figure}[t!]
  \centering
  \includegraphics[width=1.0\columnwidth]{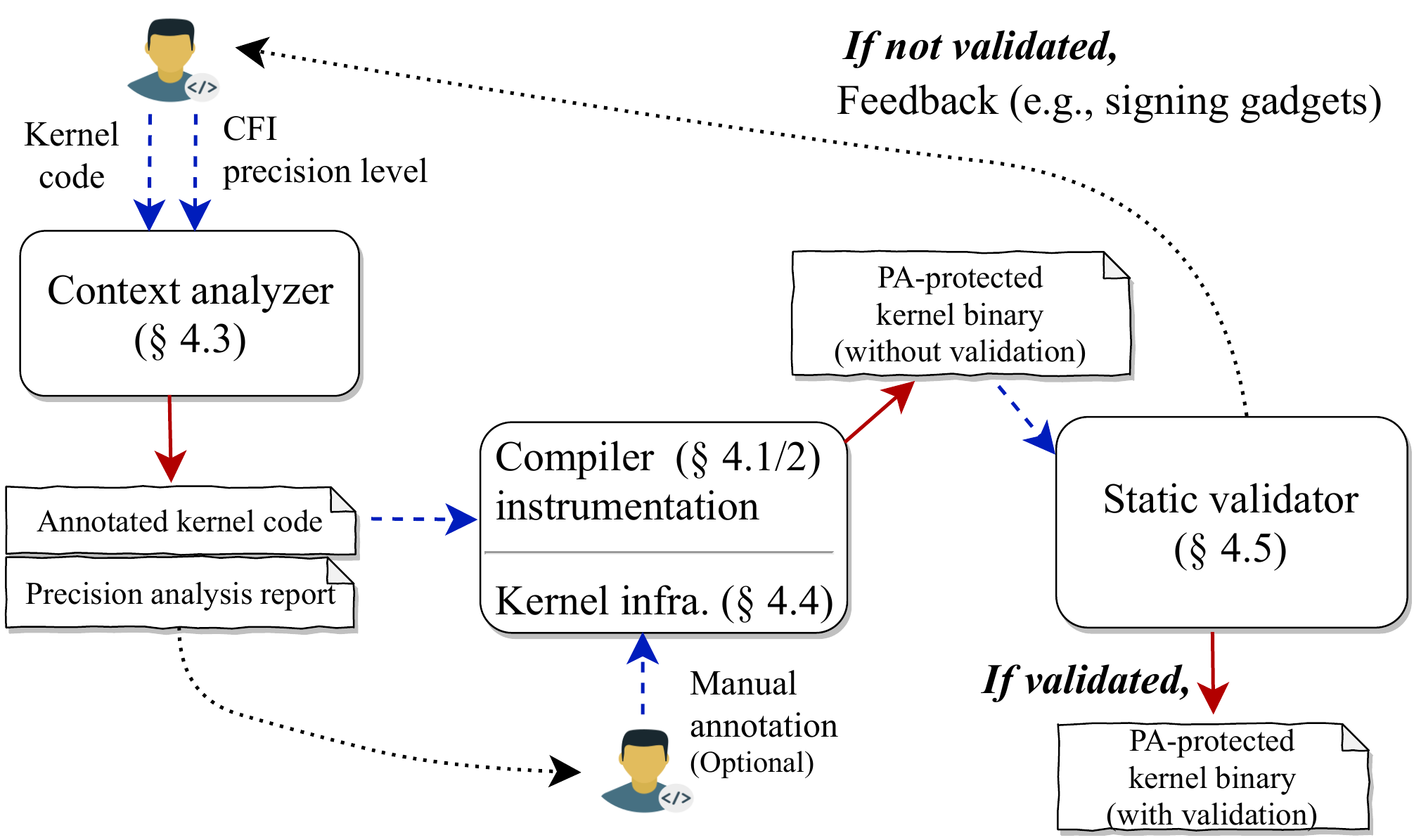}
  \caption{\sys's workflow.
    %Given a LLVM IR of kernel,
    %context analyzer provides guides for annotations.
    %Developers can selectively annotate the source code according with such guides.
    %Then given annotated code, it produces a \sys-enabled kernel.
    %Lastly, it validates that the instrumented kernel
    %respects a set of security properties
    %for the safe uses of ARM PA.
    }
  \label{f:overview}
\end{figure}

Each component of \sys
works as illustrated in~\autoref{f:overview}.
First, the context analyzer (\autoref{ss:context-analyzer})
takes kernel code and the desired CFI precision level
(i.e., the number of the allowed targets) as inputs,
and returns 1) kernel code annotated with the best run-time contexts
to meet the given precision,
and 2) a summary of the precision analysis
that describes the effectiveness of each PA context
(see, \autoref{table:allowed-target}) as outputs.
%and helps figure out remaining low-precision contexts
%even after applying 1).
%(i.e., contexts in which context analyzer cannot automatically find a way to refine up to the given precision level)
Besides that automatic annotation, developers can add annotations by hand based upon their needs.

Second,
the compiler tool-chain (\autoref{ss:compiler}, \autoref{ss:context})
performs instrumentation according to the given annotated code,
and outputs a \sys-enabled but not validated kernel binary.
%to refine low-precision contexts included in the precision analysis report.

Third, static validator (\autoref{ss:validator}) validates that the kernel binary respects a set of security properties
for the safe uses of ARM PA, and informs developers if any violation is detected.
Developers are in charge of eliminating such a violation by modifying either source code or a compiler tool-chain.

In addition, as groundwork, \sys modifies kernel infrastructure
for secure management of the protection scheme (\autoref{ss:infra}).

%context analyzer (\autoref{ss:context-analyzer}) which guides
%kernel developers how to refine PA contexts through annotations,
%a compiler tool-chain for analysis and instrumentation
%(\autoref{ss:compiler}, \autoref{ss:context}),
%a kernel infrastructure
%for secure management of the protection scheme
%(\autoref{ss:infra}),
%and a static validator
%to ensure our security properties
%of the instrumented binary
%(\autoref{ss:validator}).

\subsection{Compiler Instrumentation}
\label{ss:compiler}

To protect the integrity of function pointers,
\sys intervenes at two life-cycle events
of each function pointer:
its generation (\cc{GEN}) and use (\cc{USE}).
Simply put,
\sys signs
a function pointer at its generation
and authenticates
it right before use.
Since neither the compiler nor the PA hardware
can track life-cycle events(\cc{GEN} and \cc{USE})
of a pointer completely,
the security claim of PA-based solutions
is largely dependent on
the proper selection of a PA context,
which is bound to the pointer at \cc{GEN}
and used for authentication before \cc{USE}
(see~\autoref{ss:context}).

In this section,
we first describe our design decisions
relevant to static analysis and instrumentation:

\PP{Life-cycle of function pointers}
Function pointers in C are generated (\cc{GEN})
from a function designator~\cite{c-function-designator}---%
an expression that has a function signature
such as a function declaration or casting.
More specifically,
\sys instruments three places to insert an authentication code:
variable initialization, assign statements,
and parameters for function calls.
For each function designator,
\sys handles two kinds of scenarios:
1) \emph{constant}
where the value is emitted as part of the instruction
thus can be signed \emph{in place};
and 2) \emph{immaterial}
where the value will be determined at run-time (e.g., to support ASLR)
thus should be signed \emph{at loadtime}.
For type casting to a non-function type,
\sys leaves them intact (i.e., no authentication and re-sign),
as the dereferenced values are already signed
at their generation
and the integrity will be ultimately checked at use.
Note that such behaviors can be abused as a signing gadget by an attacker,
and our static validator is designed to identify such behaviors
(see~\autoref{ss:validator}).

An indirect call in C uses (\cc{USE})
a signed function pointer
so \sys enforces the authentication
and restores it to the original form
\emph{right before} taking it into the target function.
In addition,
\sys handles two exceptional situations
to support the practical use
of function pointers in the kernel:
function pointer comparison
and type casting to an integer.

\PP{Protecting backward edges}
\cc{GEN} and \cc{USE} of backward edges are
semantically clear: \cc{GEN} at a function call
and \cc{USE} when returning back.
To handle this case,
PA provides two dedicated instructions,
namely,
\cc{paciasp} and \cc{autiasp},
which protect a link register
that stores the address to return
after the function call
by using the address of local frame
as the context.

Unfortunately,
under \sys's strong threat model,
the current scheme is vulnerable to
a replay attack,
meaning that
an attacker can craft a signed return address
by reusing the stack memory
(i.e., repeatedly creating new tasks).
To mitigate it,
\sys simply combines a stack frame and a hash
of the current function name
as a context.
This ensures that
the signed return addresses
cannot be reused across either different functions
or different stack layouts.

\PP{Avoiding user space pointers}
% There are many active researches
% to protect user space (see~\autoref{s:relwk}).
%
\sys recognizes function pointers from user space
and selectively opts out of
signing and authentication.
In theory, it is difficult
to distinguish these two types of pointers
from a compiler's perspective,
but \sys can recognize them based on
existing idioms (e.g., \cc{__user} in Linux)
at their type declarations
and properly propagate them
in our instrumentation.
%
\begin{comment}
For example,
we verified that
\sys correctly handles
the user space signal handlers
(e.g., \cc{sigaction()} and \cc{setup\_return()} in Linux).
%
Note that protecting user pointers is
an orthogonal to what in-kernel CFI provides
and does not affect the security of kernel.
\end{comment}

\subsection{Refined Context Generation}
The granularity of protection provided by PA
depends on
the way the context parameter is generated (at \cc{GEN})
and used (at \cc{USE}).
For example, in terms of substitution attacks
(see~\autoref{s:av}),
one leaked signed pointer
can be used at any \cc{USE}
of the same context parameter---%
to be precise, the key should be equal as well,
but all pointers in the same address space (i.e., kernel space)
will be signed by the same key.

\label{ss:context}
\begin{figure}[tb]
  \include{code/obj-ex.c}
  \coderule
  \caption{Example of contexts
    based on \cc{typesig}, \cc{objtype} and \cc{objbind}
    refinement techniques
    for a \cc{IRQ} handler.
    All \cc{GEN} and \cc{USE} will
    be automatically instrumented
    given the annotation in line 5,
    automatically generated
    by the context analyzer.}
  \label{fig:obj}
\end{figure}

There are two \emph{known} techniques for
refining the context parameter further,
one using its type signature (static)~\cite{pacitup,apple-cfi}
and another using the stack pointer
(dynamic)~\cite{qualcomm-pac,apple-cfi}.
Although both approaches
can be applied without changing the original source code,
they are still far from ideal:
1) they are too coarse-grained
(e.g., one function type is used 470 times in Linux)
and 2) show high false positives
(e.g., 6.5k authentication places using \emph{zero}
as context in iOS 13).

\sys provides two static (\BC{1}, \BC{2}) and two dynamic (\WC{1}, \WC{2})
refinement techniques for context generation:

\vspace{2px}
\noindent \BC{1} \textbf{Build-time context: \cc{typesig}}.
Our \emph{baseline context} for a function pointer
is its type signature---%
a hash of its type declaration, similar to PaX RAP~\cite{pax-rap}.
With \cc{typesig} as a context in \cc{GEN} and \cc{USE},
it effectively implements
a type-based CFI
(see~\autoref{fig:obj}),
or the 1-layer confinement of MLTA~\cite{lu:typedive}.

\vspace{2px}
\noindent \BC{2} \textbf{Build-time context: \cc{objtype}}.
A type signature can be further refined
with a corresponding owner's type
when a function pointer is owned by a kernel object
(e.g., \cc{irq_handler} in \cc{irqaction} in~\autoref{fig:obj}).
For example,
one common function signature
(\cc{void (*) (void *)})
is used in 170 different indirect calls (\cc{USE})
and introduced from 200 different function designators (\cc{GEN})
in Linux.
With this refinement%
---effectively the 2-layer confinement of MLTA~\cite{lu:typedive}---%
this signature as a result
can be refined to 35 different contexts
during compilation.

\sys can further refine the context generation
to the granularity of each function pointer instance
\emph{at run-time}---%
each instance of the kernel objects
is assigned with a unique context
for protection.
The key idea is
to take advantage of
the idiomatic design patterns
used in the OS kernels,
which are commonly enforced at the code review
or as part of the maintenance cycle~\cite{linux-driver-model}.
And the idea has fully proved by \sys,
especially the context analyzer (see, \autoref{s:eval:analyzer}).
In particular,
\sys provides this scheme
as two annotations
to capture common invariants
of a relationship between a function pointer
to an object (\cc{objbind})
and to an invocation context (\cc{retbind}).
The context analyzer can automatically generate these annotations
so that developers do not need to worry about
how they supplement these annotations properly.

\vspace{2px}
\noindent \WC{1} \textbf{Run-time context: \cc{objbind}}.

\boxbeg
Annotation: \cc{objbind(\{\&\}?field, \{*|fptr\}+)}
\boxrule
This specifies which \emph{field} or its address
(\emph{\&}) should be bound
to which function pointers (one or more, or all with \cc{*})
in an object's declaration.
\boxend

\noindent
As an embedded function pointer
is often invariant
over the lifetime of its owner object,
an \cc{objbind} annotation indicates
the compiler has to bind the authenticity of the function pointer
to various properties of \cc{struct}.
Once a \cc{struct} is annotated at its declaration,
all objects of the \cc{struct}
will be instrumented
to have a dynamic context,
thus uniquely \emph{binding} the function pointer
to the created object (\cc{GEN} of the member function).
For example,
\cc{irq_handler()} can be bound
to the device's \cc{name}
(i.e., a pointer to a static string)
as in~\autoref{fig:obj},
limiting the target places for
the leaked \cc{irq_handler()}
to the one it was originally in.
Accordingly,
it can be viewed
as the 3-layer confinement
of MLTA~\cite{lu:typedive}.

\sys implements a generic technique
to composite
multiple contexts together
by \emph{chaining} the result of a previous \cc{pac}
as a context argument of the subsequent \cc{pac}
(see~\autoref{fig:obj}).
This technique,
unlike simple \cc{xor} of multiple contexts,
provides better security,
especially when an attacker
chooses an arbitrary value as one of the context
(e.g., a device name).

% Dynamic context is combined by
% \cc{pac} instructions with other contexts,
% instead of \cc{xor} like other build-time contexts use (see~\autoref{fig:obj}).
% \cc{xor} is too simple
% to prevent inferencing its result,
% makes attackers achieve
% signing gadgets
% by exploiting dynamic context.

At a glance, one would imagine
binding all the embedded function pointers
in its object's base pointer,
but this results in too many false negatives
for automation---%
for example, when an object is \cc{memcpy()}-ed,
all signed function pointers should be properly
resigned for the new context, namely,
the new object pointer
as well as its types.
This not only is fragile but incurs
high performance overhead for \cc{memcpy()},
which is commonplace in Linux (see~\autoref{t:context-change}).
%The correctness of our annotations is
%largely dependent on developers,
%yet frequent idioms or design patterns
%just need to be annotated once for all.

%The correctness of our annotations is
%largely dependent on developers,
%yet with context analyzer (\autoref{ss:context-analyzer})
%any of them can correctly annotate without detailed domain knowledge.
%

% For example,
% \cc{route_free()} of \cc{dma_router}
% is unique to a device (i.e., \cc{dev})
% as well as its type.
% %
% It's worthwhile to note that
% both objects containing
% \cc{irq_handler()} and \cc{route_free()} pointer
% are still compatible
% with \cc{memcpy()}
% unlike an object-based approach.

%% about layer-3 confinement
% \begin{figure}[t!]
% 	\include{code/pacbind-level3.c}
% 	\coderule
% 	\caption{An example of \cc{objbind}, 3-layer confinement of MLTA~\cite{lu:typedive}.}
% 	\label{fig:obj-level3}
% \end{figure}

\vspace{2px}
\noindent \WC{2} \textbf{Run-time context: \cc{retbind}}.

\boxbeg
Annotation: \cc{retbind(\{params\}+)}
\boxrule
This specifies its calling context is bound to which function arguments (\emph{params})
at the function's declaration.
\boxend

\noindent
A \cc{retbind} annotation indicates
the compiler has to bind a function pointer
to its calling contexts,
which is effective in protecting
a function pointer not embedded
to its owner object.
One such design pattern
is reference counting in Linux---%
\cc{kref} where its \cc{release()} function
is not stored as part of the object
but should be provided
together with the \cc{kref_put()} function
for reclamation (see~\autoref{fig:retbind}).
Note that this pattern saves a lot of memory uses,
as \cc{kref} instances outnumber
their invocation sites.

As \cc{kref} is frequently used in Linux
(e.g., over 110 release functions
and 127 call sites),
an attacker would substitute
any counterfeited function pointer
(e.g., via signing oracle (~\autoref{s:forging-pointers}) or leaked pointer)
to any one of such candidates.

With \cc{retbind},
a function pointer
becomes unique per calling context---%
the leaked \cc{release} pointer can be used
only at the legitimate calling context.
%
\begin{comment}
The \cc{\#call-depth} parameter refines
the call contexts even further:
it can be used for a design pattern
that has several layers of wrapper calls
to invoke a specified function.
For example,
\cc{inode_insert5()} in Linux
contains three call sites in its own calling context (depth zero)
and 33 call sites in the callee's context (depth one).
%
\end{comment}

\begin{figure}[tb]
	\include{code/retbind.c}
	\coderule
	\caption{Example of \cc{retbind}
                in protecting \cc{kref} with \sys.
	All \cc{GEN} and \cc{USE} will
    be automatically instrumented
    given the annotation in line 10,
    automatically generated
    by the context analyzer.}
	\label{fig:retbind}
\end{figure}

\subsection{Context Analyzer}
\label{ss:context-analyzer}

\sys provides a context analyzer that
spots adequate places for adopting run-time contexts
with the given kernel code and the desired precision level.
Through an inter-procedural analysis on the IR level,
the context analyzer automatically annotates
places having lower precision even with build-time contexts.
%
%It statically calculates the number of allowed targets
%for each context (\autoref{s:eval:security}),
%and suggests effective annotation schemes
%by using simple heuristics.
%
%Given the precision level the OS developers wishes to achieve,
%the suggested annotations
%can be adopted
%to protect the code pointers.
%

For \cc{objbind},
the context analyzer first estimates
each structure's \emph{diversity score},
representing each fields' compile-time diversity---%
it simply counts the number of assignments
of a new constant or a stack address or the address of heap objects.
The rationale behind choosing these as criteria is twofold--
1) a new constant intuitively means more diversity in a field value,
and 2) stack and heap address would have sufficient randomness to be used as run-time contexts
if they are newly allocated (i.e., current stack frame address or address that comes from heap allocator).

As a running example,
\autoref{fig:analyzer-objbind} shows
how to estimate the \cc{s1.p}'s diversity score.
First, the context analyzer collects all assignments
related to \cc{s1} structure (line 3, 14)
and starts dafa-flow analysis at each of them.
Specifically,
at line 3, data-flow analysis starts
on \cc{p} across function boundaries
as \cc{p} cannot be resolved within \cc{init_s1()},
so it continues using a worklist algorithm
until the value of \cc{p} is statically determined.
As a result, the diversity score will increment at line 8, 9, 19,
but not at line 10, 11.
In case that a score depends on other structure's (line 14--\cc{o1} object),
it first estimates \cc{o1.p}'s score and accumulates to the score.
Lastly, the context analyzer starts
to annotate \cc{objbind} to structures
with the highest diversity score first
until they meet the given precision level.
(see \autoref{appendix:analyzer-algo} for the more detailed algorithm)

%According to our evaluation,
%19 out of 20 \cc{struct}s having
%the most allowed targets (measured at run-time)
%rank top 10\% in diversity score.
%

For \cc{retbind}, the context analyzer identifies functions
that take a code pointer as input and consumes it in place,
then estimates the number of call sites
(e.g., for \cc{kref} (\autoref{fig:retbind}),
the function is \cc{kref_put()},
and its number of call sites is 127.)
Lastly, the context analyzer starts
to annotate \cc{retbind} to functions with all of related call sites.

%protecting the top \cc{function} bhaving the most number of call sites.
%
%Note that currently, the analyzer is not able to automatically apply \cc{retbind}
%for ones having equal or more than one call-depth, meaning that
%it only can be done by hand.

Besides automatic annotation,
the context analyzer provides a CFI precision report of the kernel
(see, \autoref{table:allowed-target}).
By analyzing the report,
developers can identify unresolved low-precision contexts
and refine them via manual annotations.

\begin{figure}[t!]
  \include{code/analyzer-objbind.c}
  \coderule
  \caption{A running example to estimate diversity score for objbind}
  \label{fig:analyzer-objbind}
\end{figure}

\subsection{Kernel Infrastructure}
\label{ss:infra}
Under our threat model
where an attacker can launch arbitrary memory reads and writes
at arbitrary moments,
the kernel should take various design decisions
into special consideration:
1) making sure that
plain function pointers are never
stored in memory
or unintentionally spilled from register files
via preemption,
2) effectively mitigating the brute-force attacks,
3) managing the keys' life-cycle
without ever storing them in memory.

\PP{Preventing preemption hijacking}
\label{s:prevent-preempt-hijacking}
To prevent TOCTOU (\autoref{s:preemption-hijacking}),
\sys should sign/authenticate the preemption context
while ensuring that it never acts
as a \emph{signing oracle} (\autoref{s:forging-pointers})
nor remains vulnerable against a \emph{replay attack} (\autoref{s:replaying-attack}).
We achieve that by introducing two new techniques as follows:
%Note that the preemption context can be created in a number of scenarios
%such as context switching routine (i.e, IRQ handler, switching between user processes),
%creating a new thread state (e.g., \cc{arm_saved_state_t} in iOS).
%this means the signing code
%of the context switching routine
%(e.g., IRQ handler, \cc{cpu_switch_to()} in linux, etc)
%never uses the attacker-chosen context.
%

\PP{1) Secure signing via key-chaining technique}
A simple approach to sign the entire preemption context is
signing each register \emph{individually}.
Unfortunately,
it is vulnerable against a replay attack
because an attacker can \emph{selectively} substitute registers
in the preemption context for control-flow hijackings.
To overcome this problem,
\sys uses the key-chaining technique
that signs each register
with the previously signed code as a context
in the chain (see, \autoref{fig:preemption}).

\PP{2) Timebind: using a timestamp as a PA context}
The above scheme prevents
attackers from modifying individual fields of the execution context,
but one can replace the \emph{whole} preemption context,
similar in concept to a replaying attack.
There are three potential defenses:
1) Similar to~\cc{typesig},
we can use a certain signature presenting the preempt context as a PA context.
However, it leaves the large number of substitution targets~\cite{mosec-pac}
since all preeempt context are signed using the same PA context;
2) Similar to~\cc{objbind},
we can use the base address of the preemption context
as a PA context.
However,
this approach still leaves
a \emph{universal} signing oracle
because an attacker would control the preemption moment
to generate a signed register value
that can be used to substitute
the same register field on the target preemption context
allocated at the same memory region (see \autoref{appendix:preempt-attack} for more detail);
3) Another solution to avoid the signing oracle
is to simply dedicate another PA key (i.e., \cc{pacga})
for signing and authenticating the preemption context,
leaving no other keys for userspace or the hypervisor.

To avoid the signing oracle without an additional PA-key,
\sys introduces a notion of \cc{timebind}
that uses an unmodifiable one-time value, \emph{timestamp}
and the base address of the preemption context,
to generate a unique context parameter for signing
(\autoref{fig:preemption}).
To get a timestamp that is resilient against the forgery,
\sys uses the Physical Timer Count on Aarch64
that monotonically increases once the system boots
but cannot be changed by system software.
This scheme prevents the replay attack
because the context will be different at every time
and it requires two additional fields (\cc{pac} and \cc{time_pac})
in the preemption context.

Note that
if an attack get to know the two additional fields,
the security of this scheme will not be negatively affected
because attacker still does not know the PA key.

\begin{figure}[tb]
	 \include{code/preemption.c}
	\coderule
	\caption{Simplified pseudo code to explain
          the use of key-chaining and timebind techniques
          to prevent preemption hijacking.}
  \label{fig:preemption}
  %% NOTE:
  %% Not possible to use autib in authenticating preemption_context
  %% due to the aspect of signing data.
  %% e.g. In the case of that attacker modify x7
  %% (1) x7 = autib(x7, x8)  // x7 changed
  %% (2) x6 = autib(x6, original-PACed-x7)
  %%     --> here we have to use original PACed-x7 for comppatiblity,
  %%     --> but then modified x7 is not going to affect the aut result.
  %% So we use pac/aut preemtion context in the way of what pacga does.
\end{figure}

\PP{Mitigating brute-force attacks}
As the number of bits allocated
for PAC is physically limited (15 bits),
an attacker can launch a brute-force attack
against the PA-protected indirect calls.
Since it is not feasible
for the kernel to simply halt the entire system
upon authentication failure,
an attacker would just
enumerate $2^{15}$ possible PACs,
which takes a few minutes (e.g., 15-min in Google's PoC~\cite{mosec-pac}).
To mitigate such a scenario,
\sys senses the forgery attempts
and backs off the execution
with a delay increasing
exponentially at every trial
based on the context;
this means an attacker-chosen context
with a randomly chosen PAC
would delay the testing oracle
exponentially.
An attacker might change the context
to bypass the back-off delay
for incorrect authentication,
but it does not increase
the likelihood of selecting the correct PAC
for the new target address.

The back-off history and strategy
need special care
to securely mitigate
the outstanding attacks---%
for example, an attacker can
try to manipulate the hashmap
that records the number of authentication failures
to render the back-off ineffective.
To address this situation,
\sys manages
the number of authentication failures
per context
on a read-only memory region
and temporarily makes it writable
to update
for a short window of time
while halting the entire machine.
This strategy not only prevents
a concurrent attack
from forging
the faulting history (as halted),
but also does not exhibit
the overhead to the normal execution
(such an event would not happen
except for under attack).
%
%\sys also guarantees that
%the fault handler never uses
%the data from memory
%that affect the back-off algorithm
%by manually auditing
%the final image.

\PP{Key management}
\label{s:key-management}
The security of PA relies on
the secrecy of its keys.
Given a leaked key,
an attacker can counterfeit
a function pointer
via cross-EL attacks
because its signing algorithm is publicly known.
In our threat model,
an attacker in user space
can forge a code pointer to jump to an arbitrary location,
say an ROP chain,
in the kernel.
Finally, the attacker can steal
data that they want via arbitrary read.

Therefore,
once the PA keys for kernel are generated at boot time,
\sys guarantee that they are never stored
into memory
during execution
to protect the key from an attacker
capable of reading an arbitrary memory region.
For key generation,
\sys leverages the randomness
provided by the bootloader
via a device tree
and utilizes the HW-based random generator if available.

Not to store the PA keys for kernel,
another important design decision is
that user spaces do not
share the same PA-keys as kernel space,
meaning the kernel and user space
have a dedicated set of PA-keys---%
B (APIB and APDB) for the kernel and
A (APIA, APDA and APGA) for the user spaces.
Moreover, at context switching,
\sys inverts all bits of the key in place
(see~\autoref{fig:keyswitching})---%
user programs can sign pointers with the \emph{inverse key}
but cannot infer the original key.
In conclusion,
the PA keys for kernel do not need
to be stored for context switching.

\begin{figure}[t!]
  \begin{minipage}{.48\columnwidth}
    \include{code/keyswitching1.s}
  \end{minipage}
  \begin{minipage}{.4\columnwidth}
    \include{code/keyswitching2.s}
  \end{minipage}

  \coderule
  \caption{\sys inverse all bits of the PA key in place
    to prevent cross-EL attack (\autoref{s:cross-el-attack})
    at context switching where preemption is disabled.}
  \label{fig:keyswitching}
\end{figure}

\sys dedicates each key to the kernel and user space,
which restricts the number of
available PA-based protection domains
to only one at a time.
To avoid this problem,
CPU designers would
consider either using independent sets of keys per execution domain,
or adding per-domain nonces
in the key assignment
of each execution level.

% \PP{Module loading}
% \sys supports module loading and unloading
% while providing the PA-based protection.
% Similar to what \sys generally treats variables,
% it signs the function pointers
% in the module object
% (e.g., \cc{init()} and \cc{exit()})
% as well as
% variables
% used in the module
% before the loading process starts.

\subsection{Static Validator}
\label{ss:validator}

The correctness of existing PA-based solutions
is largely dependent on
the correctness
of the compiler's back-end logic
like optimizations and machine-code generation.
%i
Due to complexity of whole compiler code,
it is an error-prone task
for a compiler writer
to guarantee that
PA-related concerns written at the higher layer
are preserved,
even after many stages,
at the lowest layers like produced binary.
For example, Google Project Zero
recently discovered a security hole
in iOS~\cite{google-project-zero}
where an address of a jump table
for a switch statement
was hoisted out of a for-loop
and stored in memory
due to the large number of registers used
in the for-loop.
In our threat model,
an attacker can hijack the control-flow
by crafting the stored address of the jump table
at the moment.

\sys's security, however, relies on
the correctness of the static validator,
which independently certifies that
the \emph{produced binary} respects
a set of security-critical invariants
and assumptions
taken during the compilation.
This design separation
greatly simplifies
implementation of \sys,
using a higher IR layer
(i.e., GIMPLE in the GCC)
without being concerned about the potential interference from
the back-end optimizations or machine code generation.
In addition,
our static validator
can be used to evaluate
other PA-based solutions,
such as Apple's and PARTS~\cite{pacitup}.

Our static validator
checks if four principles
that \sys assumes during the compilation
are still preserved
after the back-end optimization:

\squishlist
\item[\BC{1}] \textbf{Complete protection.}
  All indirect branches are authenticated
  and the result is checked prior to use
  (line 1 in \autoref{fig:validator-ex}).
\item[\BC{2}] \textbf{No time-of-check-time-of-use.}
  Raw pointers after authentication (\cc{aut}) or clearance (\cc{xpac})
  are never stored back in memory
  (line 6 in \autoref{fig:validator-ex}).
\item[\BC{3}] \textbf{No signing oracle.}
  There should be no gadget that signs attacker-chosen pointers
  (line 13, 18 in \autoref{fig:validator-ex}).
\item[\BC{4}] \textbf{No unchecked control-flow change.}
  All direct modifications of program counter register must be validated.
  The validator
  correctly guides us to handle special cases such as
  scheduling, signal handling, and preemption
  (see~\autoref{ss:infra}).
\squishend

% We showcase violating instances
% that the static validator
% identified for \sys as well as
% other PA-based solutions
% in~\autoref{fig:validator-p1},
% \autoref{fig:validator-p2}
% and \autoref{fig:validator-p3}.

%% \noindent
%% \BC{1}
%% \autoref{fig:validator-p1} depicts two gadgets violating \framebox{P1}.
%% In Gadget-1, the indirect branch at line5 isn't validated.
%% If an adversary could control x0 to point to her fake object address,
%% line5 could go anywhere.

%% Next, Gadget-2 from iPhone's PA CFI kernel has a dangerous instruction sequence.
%% No matter what we set to x8, the value would be always \cc{pac} properly
%% due to line8, \cc{xpac}. Consequently, line10 could go anywhere.

\begin{figure}[t!]
  \include{code/validator-ex.s}
  \coderule
  \caption{\protect\BC{1}, \protect\BC{2}, \protect\BC{3} violations from PARTS~\cite{pacitup} and iOS and \sys.
  An attacker can break PA with an attacker-chosen pointer by controlling
  either function arguments or spilled stack memory.
}
  \label{fig:validator-ex}
\end{figure}

\PP{Algorithm.}
It performs a simple \emph{intra-procedural} analysis
with CFG recovery and loop detection
given a binary image.
It first scans all instructions
within a function
and runs \textsc{validate_bb} (\autoref{a:validator-algo})
on PA instructions.

With \autoref{fig:validator-ex} as a running example,
we first describe the cases in which the analyzer can detect violations within a basic block.
To validate \BC{1} and \BC{3},
it invokes \textsc{validate_bb}
on \cc{blr} and \cc{pacia}, respectively (line 4, 16 in \autoref{fig:validator-ex})
with \cc{x21} as a symbolic register $sym$
in~\autoref{a:validator-algo}.
Then, \textsc{validate_bb} attempts
to find the origin of \cc{x21}
in a backward recursive way
by exploring all possible paths.
Conservatively,
due to \cc{ldr} (line 3, 15 in \autoref{fig:validator-ex})
in a previous path,
it reports a violation (line 4 in \autoref{a:validator-algo}).
If $sym$ is originated from the function parameters (line 19 in \autoref{fig:validator-ex}),
\textsc{validate_bb} would conclude
as a potential violation
as it cannot be resolved
even after exploring the whole function.
To validate \BC{2},
it starts from \cc{autib} (line 7 in \autoref{fig:validator-ex}) with \cc{x2} as $sym$
and attempts to find the uses of \cc{x2} in a forward recursive way,
and then reports a violation because of \cc{stp} (line 8 in \autoref{fig:validator-ex}).
As a special case, to detect the violation at line 22 in \autoref{fig:validator-ex},
it checks if a call instruction places between address calculation (line 23) and PA instruction (line 25).
This trick enables detecting such a violation without inter-procedural analysis but entails false positives
because a register containing PA-relevant values might not be stored in the memory.
%because the validator cannot guarantee the register containing PA-relevant value is stored onto the memory

The validator, of course, can work across basic blocks in the following cases--
1) if $sym$ cannot be resolved within a basic block (i.e., reaching line 18 in \autoref{a:validator-algo}), the validator recursively invokes \textsc{validate_bb}
on all predecessors of the current $bb$ (line 21 in \autoref{a:validator-algo}),
and 2) If we encounter any of branch instructions jumping to somewhere in the current function before $sym$ is resolved,
it invokes \textsc{validate_bb} on the target basic block. (line 17 in \autoref{a:validator-algo})

\PP{Results.}
%\XXX{any problem of this?}
We applied static validator to PARTS, iOS kernel and \sys itself,
as a result, confirmed 15/5/0 violations respectively.
We found 7 violations during \sys development,
1/1/5 for \BC{1}/\BC{2}/\BC{3} respectively,
and fixed all by modifying either our compiler pass or
kernel code.
%\XXX{add bugs in \sys (shortly)}.
% Note that we excluded false positives with manual efforts
% to get the refined numbers.

\PP{False positives.}
Since it is infeasible to implement perfect binary analysis on the kernel,
the validator reported about 100 false positives, and their root causes were mainly due to--
1) too complicated control flows in the kernel, meaning that too many basic blocks are involved in a PA instruction,
or 2) the absence of inter-procedural analysis that is needed to detect the violation at line 22 in \autoref{fig:validator-ex},
or 3) uncertainty of the type of memory to be loaded (e.g., if \cc{x19} at line 15 in \autoref{fig:validator-ex} points to read-only memory, that is not the violation).

The task to eliminate those took two days by a person who is knowledgeable with binary analysis.
We plan to improve our static validator to reduce such false alarms as future work.

\SetNlSty{}{}{}
\let\oldnl\nl% Store \nl in \oldnl
\newcommand\nonl{%
\renewcommand{\nl}{\let\nl\oldnl}}% Remove line number for one line
\SetAlCapNameFnt{\small}
\SetAlCapFnt{\small}

\begin{figure}
\footnotesize
\SetKwFunction{validate}{\textsc{validate_bb}}
\SetKwInOut{Symbol}{Symbol}
\SetKwInOut{Input}{Input}
\SetKwInOut{Output}{Output}
\SetKwProg{ValidateBB}{\textsc{validate_bb}}{}{}
\nonl\ValidateBB{($bb,si,sym$)}{
  \Input{$bb$: a basic block to be inspected\newline
  $si$: a first instruction to be inspected in $bb$\newline
  $sym$: a symbolic register containing PA-relevant value}
  \Output{true if no violations, otherwise false}
  \Symbol{
  $A$: arithmetic/bitwise instructions\;
  $L$: load instructions\;\newline
  $C$: address calculation instructions\;
  $P$: predecessors of $bb$\;\newline
  $i_{*op}$: source/destination register operand of $i$\;}
  \vspace{2px}
  \For{$i \gets si;\ i \neq bb.first();\ i \gets i.prev();$}{
    \If{$i \in A \And i_{destop} = sym$}{
      $sym \gets i_{srcop}$\;
    }
    \ElseIf{$i \in L \And i_{destop} = sym$}{
      \Return false;
    }
    \ElseIf{$i \in C \And i_{destop} = sym$}{
      \Return true;
    }
    \ElseIf{$i = \text{"auti*"} \And i_{destop} = sym$}{
      \Return true;
    }
    \ElseIf{$i = \text{"xpaci*"} \And i_{destop} = sym$}{
      \Return false;
    }

    // call instruction

    \ElseIf{$i = \text{"bl"}$}{ 
      \Return false;
    }

    // jump or conditional branch instruction

    \ElseIf{$i = \text{"b"}$}{
      $target \gets i.target$\;

      \Return \textsc{validate_bb($target,target.end(),sym$)};
    }
  }
  \If{$P = \varnothing$} {
    \Return false;
  }
  \ForEach{$bb \in P$}{%
    \If{not \textsc{validate_bb}({$bb,bb.end(),sym$})}{%
      \Return false;
    }
  }
  \Return true;
}
\caption{
  The core algorithm to verify \protect\BC{1}, \protect\BC{3} and \protect\BC{4}.
  The algorithm for \protect\BC{2} is explained in \autoref{ss:validator}.
}
\label{a:validator-algo}
\end{figure}

%\PP{Miscellaneous things}

%- pacing ro/global pointers, jmptables

%- fault/bruteforce detection $\rightarrow$ fail-stop (brk 1 vs *acc)

%- JIT/BPF support

%- Return address signing w/ kernel stack randomization

\section{Implementation}
\label{s:impl}

We implemented \sys's instrumentation
by adding a new GIMPLE pass (after the CFG pass)
on the GCC 7.4.0 as a plugin.
Also, we put very small modification to the GCC,
in order to interpose on function prologues and epilogues
for the backward-edge CFI.
We chose the GCC as a first target,
to actually deploy the CFI protection
to our commercial products,
such as appliances, IoT devices, and smart phones,
that rely only on the GCC.
We also implemented instrumentation in LLVM, but it supports only a subset of \sys features.
(e.g., \cc{typesig}).

\begin{table}[t!]
\centering
\scriptsize
\begin{tabular}{lr@{ }l}
  \toprule
  Components & \multicolumn{2}{l}{Lines of code} \\
  \midrule
  the GCC plugin         & 3,632 & LoC (for forward-edge CFI)\\
  the GCC                & 127   & LoC changes (for backward-edge CFI)\\
  Linux              & 491   & LoC changes \\
  \quad \cc{objbind} &  87   & structs \\
  \quad \cc{retbind} &  79   & functions \\
  FreeBSD            & 258   & LoC changes \\
  \quad \cc{objbind} &  25   & structs \\
  \quad \cc{retbind} &   3   & functions \\
  Static validator   & 848   & LoC (python)\\
  Context analyzer   & 1943   & LoC (c++)\\
  \bottomrule
\end{tabular}
\caption{The complexity of \sys's components (in LoC).}
\label{t:loc}
\end{table}

\PP{Kernel modifications}
We made minimal changes to Linux
(491 LoC)
and FreeBSD (258 LoC).
We manually fixed incorrect declarations of function types
(e.g., dummy console and filler)
similar to Android's patches
to support CFI~\cite{android-cfi, dummycon-commit}.
The context analyzer automatically adds
166 annotations to Linux
and 28 annotations to FreeBSD,
for contexts having more than 100 targets even when objtype is used (see~\autoref{table:allowed-target}).
(see~\autoref{t:loc})

\PP{Preemption hijacking protection}
We prevent preemption hijacking in two places in Linux:
1) \cc{el1_irq()} called when an IRQ occurs at the kernel mode, and
2) \cc{el0_irq()} called when an IRQ occurs at the user mode.
In 1), we sign and authenticate not only all general-purpose registers (i.e., \cc{x0}--\cc{x30})
but also some special-purpose registers (e.g., \cc{elr_el1}, \cc{spsr_el1})
as in~\autoref{fig:preemption}.
%XXX:Comment{update figure5; we must sign special purpose registers for el1_irq}
%
In 2),
we simply perform sanity checks on special purpose registers
to prevent the hijacking to kernel space
instead of returning back to user space.
This protection cannot be exploited as a signing oracle (\autoref{s:forging-pointers})
because arm64 guarantees that
all registers are preserved when an interrupt is raised
and Linux does not allow nested interrupts
on both IRQ handlers mentioned above.

\PP{Backward-edge protection}
In \sys,
creating a context for backward edge protection
requires an operation with a constant
and stack pointer register(\cc{sp})
that is not allowed direct uses
as an operand in Aarch64.

For this reason,
function prologues (left-side) and epilogues (right-side) use
two additional registers---%
a register as an operand of combine instruction (\cc{bfi})
and a register as a context for PA---%
as follows.

\vspace{4px}
\begin{minipage}{.48\columnwidth}
\begin{Verbatim}[commandchars=\\\{\},codes={\catcode`\$=3\catcode`\^=7\catcode`\_=8}]
\PY{n+nf}{mov} \PY{n+no}{x9}\PY{p}{,} \PY{n+no}{sp}
\PY{n+nf}{mov} \PY{n+no}{x10}\PY{p}{,} \PY{n+no}{hash}\PY{p}{(}\PY{n+no}{FUNC\PYZus{}NAME}\PY{p}{)}
\PY{n+nf}{bfi} \PY{n+no}{x9}\PY{p}{,} \PY{n+no}{x10}\PY{p}{,} \PY{l+m+mi}{32}\PY{p}{,} \PY{l+m+mi}{32}
\PY{n+nf}{pacib} \PY{n+no}{lr}\PY{p}{,} \PY{n+no}{x9}
\end{Verbatim}

\end{minipage}
\begin{minipage}{.4\columnwidth}
\begin{Verbatim}[commandchars=\\\{\},codes={\catcode`\$=3\catcode`\^=7\catcode`\_=8}]
\PY{n+nf}{mov} \PY{n+no}{x9}\PY{p}{,} \PY{n+no}{sp}
\PY{n+nf}{mov} \PY{n+no}{x10}\PY{p}{,} \PY{n+no}{hash}\PY{p}{(}\PY{n+no}{FUNC\PYZus{}NAME}\PY{p}{)}
\PY{n+nf}{bfi} \PY{n+no}{x9}\PY{p}{,} \PY{n+no}{x10}\PY{p}{,} \PY{l+m+mi}{32}\PY{p}{,} \PY{l+m+mi}{32}
\PY{n+nf}{autib} \PY{n+no}{lr}\PY{p}{,} \PY{n+no}{x9}
\PY{n+nf}{ret}
\end{Verbatim}

\end{minipage}
\vspace{4px}

Note that those registers should be caller-saved registers (e.g., x9, x10)
to protect register spilling causing performance overhead.

\PP{Supporting Linux}
We found developer guides that motivate to devise \cc{objbind}.
Linux kernel has provided
design patterns for inside components
(e.g., device driver~\cite{linux-driver-model}),
strongly recommending developer to
use special functions and structures.
As a result, most code consists of some patterns,
which helps the context analyzer refine contexts easier
to reduce allowed targets in Linux.

\PP{Supporting FreeBSD}
We found two interesting function types---%
\cc{kobjop_t} and \cc{sy_call_t}---%
used for better software abstraction.
In other words, function pointers are stored
as different type with the actual type of
pointed function.

Finally, we found 125 and 342 function types,
stored after type-converted as \cc{kobjop_t} and \cc{sy_call_t} respectively,
which means that FreeBSD allows many allowed targets.
In \sys, the context analyzer automatically applied \cc{objbind} to refine these function types.

\PP{Context Analyzer}
The context analyzer, written in C++,
first takes as input kernel codes and builds the kernel.
Afterward, it extracts an LLVM bitcode file for the fully linked kernel binary (e.g., vmlinux for Linux)
to enable inter-procedural analysis for a whole,
and starts the static analysis.

\section{Evaluation}
\label{s:eval}

In this section,
we evaluate \sys's approach in four key areas:

\squishlist
\item[Q1.] How does our approach
  compare with known PA-based CFI solutions?
  (\autoref{s:eval:security})
\item[Q2.] How do we validate its security guarantee
  and the correct functionality of \sys-enabled kernels?
  (\autoref{s:eval:correctness})
\item[Q3.] How much performance overhead does \sys impose
  on user applications and the kernel?
  (\autoref{s:eval:performance})
\item[Q4.] How do we check the soundness and effectiveness of our context analyzer?
  (\autoref{s:eval:analyzer})
\squishend

\PP{Experimental setup}
We selected two target devices, the Mac mini (the M1) and the Raspberry pi 3,
to represent a high-end and a low-end ARM device respectively.
We applied \sys
to Linux (Asahi Linux~\cite{asahi}
customized for the M1 chip based on Linux 5.12.0-rc1,
and Linux 4.19.49 for Tizen 5.5)
and FreeBSD (FreeBSD 11.0-CURRENT),
and evaluated them on two real devices and one virtual platform:
the Mac mini for Asahi Linux and
the Raspberry Pi 3 for Tizen 5.5
and QEMU for FreeBSD.
We reported the real performance on the M1 (using actual PA instructions)
and estimated the PA's performance by measuring real cycles
taken to execute each PA instruction on both Apple A12 and the M1 on userspace
(see~\autoref{s:eval:performance}).
To provide a realistic kernel configuration,
we adopted the union of Asahi's and \cc{arm64}'s
default config for Linux 5.12.0-rc1.
Also, we used the default configs
for Tizen 5.5 and FreeBSD.

\subsection{Comparing with Other Approaches}
\label{s:eval:security}

\begin{table}[t!]
  \centering
  \scriptsize

  \centering
  \vspace{-3pt}
	\includegraphics[page=1, width=0.95\columnwidth]{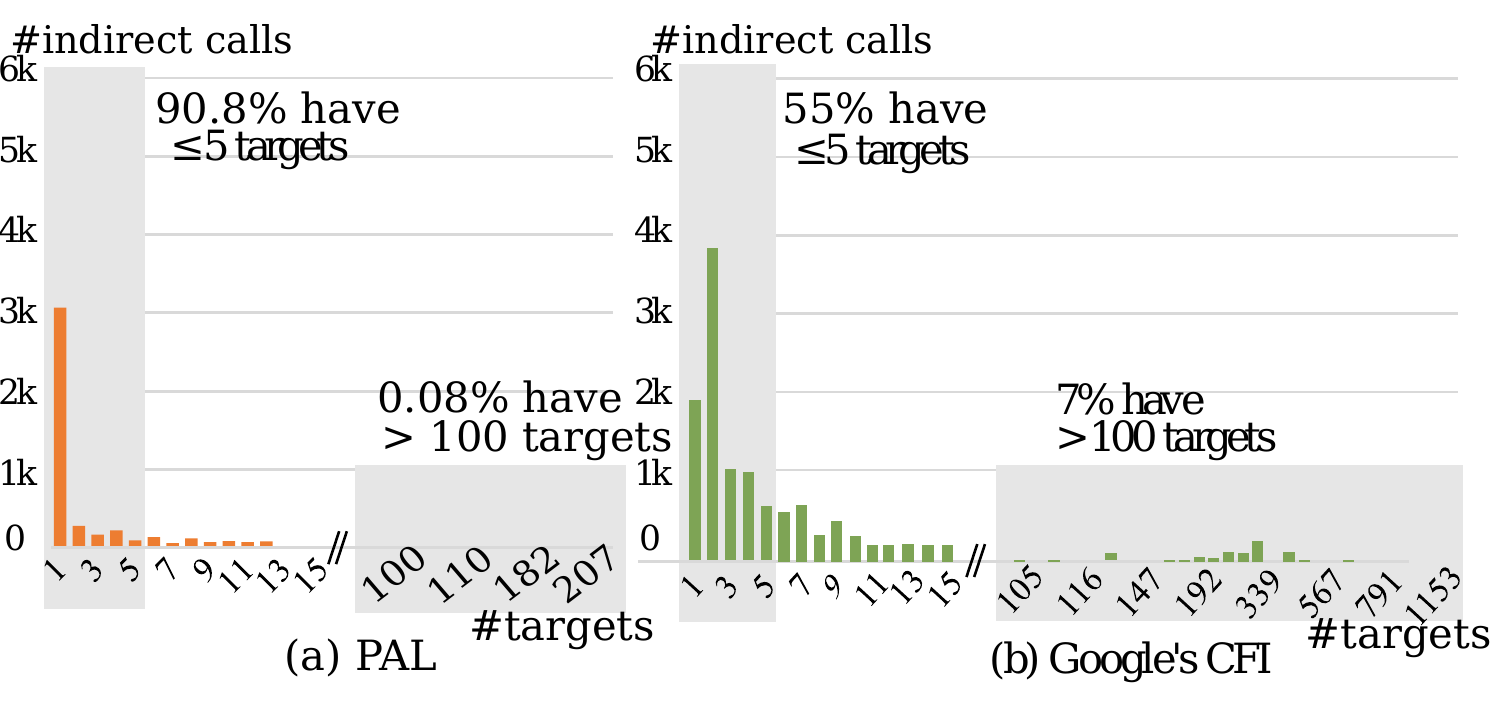}

	\begin{tabular}{lrrrrr}
 		\toprule
		\multirow{3}{*}{\#Tgts}
			& \multicolumn{1}{c}{\multirow{3}{*}{\textbf{Google's}}}
			& \multicolumn{1}{c}{\multirow{3}{*}{\textbf{BTI}}}
			& \multicolumn{3}{c}{\textbf{Linux w/ \sys}} \\
                        \cmidrule(lr){4-6} \\
		& {} & {}
			& \multirow{2}{*}{\cc{typesig}}
			& \multirow{2}{*}{+\cc{objtype}}
			& {+\cc{objbind}} \\
		& & & & & {+\cc{retbind}} \\
		\midrule
		$\leq 5 $ & $55.0\%$ & $0.0\%$ & $84.9\%$ & $88.6\%$ & $90.8\%$ \\
		$> 100 $  & $ 7.0\%$ & $100.0\%$ & $ 2.8\%$ & $ 1.6\%$ & $0.08\%$ \\
		\midrule
			Max &  1,153  & 59,300 &     35,264 &     30,622 &    207 \\
		\bottomrule
	\end{tabular}
	\caption{The precision of \sys
          in terms of the number
          of \emph{allowed} indirect call targets.
          We show $\leq 5$ and $> 100$
          in comparison with reported Google's CFI
          applied to Android 4.14~\cite{androidblog-cfi}.
          We also added ARMv8.5 BTI scheme for comparison.
          It shows \sys can effectively enhance the precision
		  of the in-kernel CFI.
		  }
  \label{f:allowed-target}
	\label{table:allowed-target}
\end{table}

We first compare the precision of \sys's protection
with two other state-of-the-art CFI schemes
that have been deployed on Android.
Then, we compare ours
with two other PA-based protection schemes,
namely, PARTS~\cite{pacitup},
and iOS's CFI (since iPhone XS).
Last, we compare PA-based solutions
in terms of PA context changes.

\PP{Allowed targets for indirect calls}
The precision of forward-edge CFI
can be estimated
by counting the number of
allowed targets for each indirect call.
% (all backward edges are also protected in \sys).
%
In PA,
an indirect call can be taken to
any locations
if the function pointer
is signed by the same context
used in the call site.
Since our static validator checked that
there is no signing gadget embedded in the final binary,
the precision can be measured
by simply counting the number of pointers
signed by the same context.
\autoref{table:allowed-target}
shows the total number of allowed targets
by each indirect call---%
if there are two call sites (\cc{AUT}) and five different calls (\cc{PAC}) using the same context,
we conservatively estimated its allowed set to 10.
We ran the context analyzer with 100 allowed targets as the precision level
and used the automatically annotated kernel code that the analyzer produced.

Compared
to the state-of-the-art CFI protection applied to
Android~\cite{androidblog-cfi},
\sys improves the precision of CFI
significantly:
the number of indirect calls
with fewer
than five targets
increases by 5.9\% (to 90.8\%)
and the ones with more than 100 targets
decreases from 2.8\% to 0.08\%.
Most important, only 3 contexts are included in $> 100$ after applying both objbind and retbind.

We reference estimations from
Google's public report on
the precision of the deployed CFI
on recent Android~\cite{androidblog-cfi}.
The differences
in Google's type-based CFI and our \cc{typesig}
are due to
the version differences
of each kernel-- 4.14 in Google's
and 5.12.0-rc1 in \sys--
as well as \sys's large kernel configuration.
As a comparison, we also added the estimation
of ARM's hardware-based CFI,
BTI (Branch Target Identification) introduced in ARMv8.5~\cite{bti},
which only limits all indirect transitions
to the function entry.

\autoref{table:allowed-target} also shows
the effectiveness of refined context generation
used in \sys.
Our static context (i.e., \cc{objtype})
reduces the overall number of target sets
while the run-time contexts (i.e., \cc{objbind} and \cc{retbind})
effectively refine
the most common call targets (from 30622 to 207).

\PP{Context diversity on indirect calls}
\label{s:context-diversity}
To compare with other PA-based solutions,
we measured the CFI precision
by counting the number of indirect calls
sharing the same context
(shown~\autoref{table:context-diversify}).
For fair comparison,
we applied PARTS to Linux 5.12.0-rc1
and performed binary analysis
on the latest iOS firmware image.
% downloaded from t(iPhone12,3,iPhone12,5_13.0_17A577_Restore.ipsw).
Compared to PARTS,
\sys improves the small set measure ($\le 5$ contexts) from
20.7\% to 94.9\%,
and reduces the large set measure ($ >100$) from 18.4\%
to near zero.
Compared to a dynamic, kernel space protection, iOS\footnote{
  We treated \emph{dynamic} contexts as an \emph{unique} context,
  so counted in $\le 5$.
},
\sys also effectively refines the attack targets:
it reduces the large set measure from 21.2\% to near none
while eliminating using the zero context
(i.e., 6513 indirect calls using the zero context in iOS).
Note that iOS's the large set measure (21.2\%) is due to
not only the zero context but also
the context containing offsets for jump tables.

\PP{Context changes}
\label{s:context-changes}
PA-based solutions are often required to change
the context used to sign a pointer:
e.g., type-casting on a function pointer requires
authentication with a previous context
and re-signing with a new context.
For PA solutions relying on \emph{static} contexts,
this task is straightforward,
but for ones using \emph{dynamic} contexts,
this conversion is often implicit and non-trivial to handle
(see~\autoref{t:context-change}).
For example,
when an object is copied with \cc{memcpy()} or \cc{memmove()},
the member functions are no longer considered
properly signed
with the context (e.g., the base address of the object).

For this reason,
iOS uses \emph{zero} context for all C function pointers
as well as C++ V-Table pointers (not entries inside the table)~\cite{ios-mac-security},
which can be vulnerable to replay attack as demonstrated by Google project zero.~\cite{google-zero-context}
Meanwhile, PATTER~\cite{patter} interposes these memory-related functions,
checks each byte of source to identify a signed pointer,
and re-signs with a new address,
which can be leveraged for a signing gadget (\autoref{s:forging-pointers})
because it ignores authentication failures.
In contrast,
\sys's approach capturing the kernel's design patterns
(see~\autoref{ss:context})
performs in a robust manner.

\begin{table}[t!]
	\centering
	\scriptsize
	\begin{tabular}{lccc}
 		\toprule
		\#Contexts & \textbf{PARTS} & \textbf{iOS kernel} & \textbf{\sys} \\
 		\midrule
		$\leq 5 $ & $20.7\%$ & $62.2\%$ & $94.9\%$ \\
		 $> 100 $ & $18.4\%$ & $21.2\%$ & $0.0\% $ \\
 		\midrule
		 Max &      353 &     6,513 & 70 \\
 		\bottomrule
	\end{tabular}
	\caption{The diversity of context used in \sys, iOS and PARTS
			in terms of the number of indirect calls that share equal context.
			% We show $\leq 5$ and $> 100$
			% in comparison with PARTS\cite{pacitup} and iOS kernel.
			% It shows \sys enough diversified contexts,
			% which means \sys fully use context scheme of PA.
        }
	\label{table:context-diversify}
\end{table}

\begin{table}[t!]
	\centering
	\scriptsize
	\begin{tabular}{@{}l ccccc@{}}
		\toprule
		\textbf{Context} & \multirow{2}{*}{\textbf{PARTS}} & \multicolumn{2}{c}{\textbf{iOS kernel}} & \multirow{2}{*}{\textbf{PATTER}} & \multirow{2}{*}{\textbf{\sys}} \\
		\quad property					&                                 & \textbf{C}       & \textbf{C++}      &                                  &                               \\
		\midrule
		\textbf{Build-time} & typesig & zero & namesig &  & typesig, objtype \\
		\quad compatibility & \ding{53} & \ding{108} & \ding{108} & & \ding{108} \\
		\quad security & \ding{53} & \ding{53} & \ding{108} & & \ding{108} \\
		\midrule
		\textbf{Run-time} & & & address & address & field \\
		\quad compatibility & & & \ding{108} & \ding{108} & \ding{108} \\
		\quad security & & & \ding{108} & \ding{53} & \ding{108} \\
		\bottomrule
	\end{tabular}
	\caption{
	  Comparing PA-based forward-edge protections in terms of context changes.
	  \emph{address} means the storage address of a function pointer to be signed/authenticated;
	  \emph{namesig} means a hash of mangled function name used in iOS for C++ V-Table entries~\cite{ios-mac-security};
	  \sys's approach is not only compatible with memcpy()
	  but also provides finer protection by utilizing both static and dynamic contexts.}
	\label{t:context-change}
\end{table}

% \tablename{\ref{table:context-diversify}} shows how many indirect calls that
% use the same context as others, by each binary. As a result, in the case of
% PARTS, the number of indirect calls using nearly unique contexts is the lowest,
% which means that the contexts in PARTS were not enough diversified.
% Then, in the case of XNU CFI kernel, the number of indirect calls using very
% common contexts is very highest, which means that the attack surface is
% very large for forgery attacks.

% On the other hand, Our study does not have indirect calls using the same context
% as more than 100 other indirect calls. And, about three quarters indirect
% calls have enough diversified contexts.

\PP{Backward-edge protection}
\label{s:compare-backward}
Unlike Apple's primitive backward-edge protection~\cite{ios-mac-security},
\sys enhances its precision
by combining the hash of a function name
and a stack pointer,
similar to PARTS~\cite{pacitup}.
To quantify the improvement,
we estimated the maximum number
of allowed targets for backward edges
while running LMbench on Linux.
Our evaluation shows that
it effectively reduces
the allowed targets from 203 (Apple's) to 14.
%---%
%both are in \cc{_raw_spin_unlock()}.

Other finer-grained solutions
like PACStack~\cite{pacstack} or Camouflage~\cite{camouflage}
impose undesirable performance overheads:
PACStack requires excessive memory accesses
to trace every call stack~\cite{pacstack}
and Camouflage needs to reserve a register
to retain the function address
until the function epilogue~\cite{camouflage}.

\begin{comment}
\begin{table}[t!]
	\centering
	\scriptsize
	\begin{tabular}{@{}l cccc@{}}
		\toprule
		\textbf{Property} & \textbf{PA primitive} & \textbf{PACStack} & \textbf{Camouflage} & \textbf{\sys, PARTS} \\
		\midrule
		\multirow{3}{*}{Context} & \multirow{3}{*}{\cc{sp}} & chain of & \cc{sp} $\oplus$ & \cc{sp} $\oplus$ \\
		& & return& function & hash of\\
		& & addresses & address & function name\\
		\midrule
		Maximum & \multirow{2}{*}{203} & \multirow{2}{*}{1} & \multirow{2}{*}{14} & \multirow{2}{*}{14} \\
		allowed targets & & & & \\
		\midrule
		Memory access & \ding{53} & \ding{108} & \ding{108} & \ding{53} \\
%		\midrule
%		\textbf{Resistibility} \\
%		\quad{replaying attack} & \ding{53} & \ding{108} & \ding{108} & \ding{108} \\
%		\quad{arbitrary RWs} & \ding{53} & \ding{108} & \ding{53} & \ding{108} \\
		\bottomrule
	\end{tabular}
	\caption{
	  Comparing PA-based solutions in terms of backward edge protection.
	\sys's approach does not access any memory spaces, but provides
	reasonable precision of allowed targets.}
	\label{t:backward-protection}
\end{table}
\end{comment}

\subsection{Security and Functional Validation}
\label{s:eval:correctness}

\PP{Correctness testing}
We tested the correctness of the \sys-protected Linux
by applying micro- and macro-benchmarks:
LMbench, perf bench, Apache bench, leveldb, Blogbench
and Linux Test Project (LTP).
We also confirmed
that the original kernel exhibits the same behaviors
in all benchmarks.

\PP{CVE studies}
\label{s:eval:cve}
We tested three known CVEs
(CVE-2017-7308 and CVE-2018-9568 for Linux,
CVE-2019-5602 for FreeBSD)
and corresponding exploits against
both original and protected kernels.
We confirmed that all exploits are successfully prevented---%
original exploits were simply
prevented by \cc{typesig}.
However, a stronger adversary
could easily launch
a replaying attack (\autoref{s:replaying-attack}),
e.g., CVE-2019-5602 that abuses \cc{struct sysent.sy_call}
having 547 allowed targets,
which \cc{objbind} in \sys could effectively prevent.

%\begin{table}[t!]
%	\centering
%	\scriptsize
%	\begin{tabular}{lccc}
% 		\toprule
%		\textbf{CVE} & \textbf{OS} & \textbf{Context} & \textbf{Blocked}\\
%		\midrule
%		 CVE-2017-7308 & Linux & typesig+objtype+objbind & \checkmark\\
%		 CVE-2018-9568 & linux & typesig+objtype+objbind & \checkmark\\
%		 CVE-2019-5602 & FreeBSD & typesig+objtype+objbind & \checkmark\\
%		\bottomrule
%	\end{tabular}
%	\caption{CVE studies over multiple OSes.
%			We have confirmed that both build-time and run-time context
%			could be applied to the exploits on these CVEs
%			to block to corrupt their target function pointers.}
%	\label{table:cve-study}
%\end{table}

\PP{Run-time validation}
\label{s:runtime-validator}
To check if there are any \emph{overlooked} function pointers
not sanitized by our analysis,
we took a series of memory snapshots
of the running kernel while executing LMbench tests.
With this run-time validation,
we found several, non-trivial bugs
during the development of \sys:
e.g., a raw function pointer made in \cc{kernel_thread},
saved in the \cc{x19} field of \cc{task} structure,
eventually loaded and called in \emph{assembly code}
without \cc{aut} instruction.
We manually added PA instructions
to protect the function pointers.
\begin{comment}
We observed five classes of
raw function pointers stored in the kernel memory,
but \emph{without} security implications:

\squishlist
\item[1)] Return addresses of functions
  used during initialization,
  but never used after.
\item[2)] Function pointers in callee-saved registers
  spilled during the function call,
  but not used after.
\item[3)] Function pointer tables used in assembly code as jump tables
in non-writable region.
\item[4)] Non-used static constant variables
  because of compiler optimizations
  that change an indirect call to a direct call,
  yet not eliminated because of its side-effects.
\item[5)] Saved CPU contexts due to preemption such as IRQ,
  but all properly protected by \sys
  (\autoref{s:prevent-preempt-hijacking}).
\squishend
\end{comment}

\subsection{Performance Overhead}
\label{s:eval:performance}

We measured
the performance overhead
imposed by \sys
in terms of computation throughput and latency.
(see~\autoref{appendix:evaluation} for detailed numbers)
%
%Since there is no publicly available development board,
%we first explain
%how we faithfully follow the performance model
%of ARM's PA on iPhone (A12) and the Mac mini (the M1).

\newcommand{\PPP}[1]{\round{4}{#1}}

\begin{table}[!t]
       \scriptsize
       \centering
       \begin{tabular}{l c c}
               \toprule
	       \multicolumn{1}{c}{Instruction} &
               \multicolumn{1}{c}{\textbf{iPhone (A12): Time (ns)}} &
               \multicolumn{1}{c}{\textbf{the Mac mini (the M1): Time (ns)}} \\
               \midrule
\cc{paciza} / \cc{pacizb}
  & \PPP{2.817218} / \PPP{2.816156}
  & {--} / \PPP{2.150117} \\

\cc{pacia} / \cc{pacib}
  & \PPP{2.810315} / \PPP{2.817032}
  & {--} / \PPP{2.150295} \\

\cc{autiza} / \cc{autizb}
  & \PPP{2.816489} / \PPP{2.816998}
  & {--} / \PPP{2.150236}\\

\cc{autia} / \cc{xpaci}
  & \PPP{2.817679} / \PPP{2.818857}
  & {--} / \PPP{2.150422} \\
\midrule
eor/orr
 & \PPP{0.403193} / \PPP{0.402968}
 & \PPP{0.307748} / \PPP{0.307416}
\\
               \bottomrule
       \end{tabular}
       \caption{Performance of PA instructions measured on
         iPhone (A12) and the Mac mini (the M1).
         We replace each \cc{pac}/\cc{aut} instructions
         with seven \cc{eor} instructions
         to emulate the PA's performance overhead.}
       \label{t:realcycle}
\end{table}

\PP{Micro-benchmark}
We used \emph{two} micro-benchmarks:
LMbench and perf on the Mac mini and the Raspberry pi 3.

\vspace{2pt}
\noindent
\WC{1} \textbf{LMbench}.
We ran LMbench v3 to measure the potential impact
of system call latency increased by \sys.
Compared to stock Linux,
\sys increases the latency by 0-3 $\mu$s
depending on system calls, % (see~\autoref{f:lmbench}).
on both the Mac mini and Pi 3.
Due to the additional \cc{pac}/\cc{aut} instructions used
for signing and validating the process context,
the latency of \cc{fork()}
is impacted the most: 28.0~$\mu$s on Pi 3,
6.1~$\mu$s on the Mac mini.
Also, we measured the only impact of backward-edge protection,
\sys increases the latency by 0-1~$\mu$s on average.
% and \cc{fork()} and \cc{exit()} are impacted the most: 3.9~$\mu$s.

\noindent
\WC{2} \textbf{perf}.
To measure the performance overhead
associated with context switching
(i.e., the chained signing operations with the time stamp),
we ran the perf benchmark
(5.12.0-rc1)~\cite{linux-perf}.
Our experiments show
that
it degrades the latency of messaging and pipe
by 3-5\% on both the Mac mini and Pi 3.
%
%This experiment showcases
%the worst-case performance overhead
%to \sys.

\PP{Macro-benchmark}
We ran Apache benchmark (v2.3) and leveldb (v1.22) and Blogbench (v1.1)
to estimate the performance impact of
network and database and file server applications
on \sys-enabled Linux, respectively.

\vspace{2pt}
\noindent
\WC{1} \textbf{Apache benchmark}.
We used two pi 3 devices--
the client sends GET requests
with varying sizes through ethernet
to the \sys-enabled and stock Linux (server)
for comparison.
\sys degrades the performance about 1\%
for 1~KB files
but has negligible overhead (< 0.08\%)
for requests over 100~KB.

On the Mac mini,
we set up both client and server in one mac mini
due to the lack of ethernet support of Asahi linux, and
\sys degrades the performance by 0.75\% for 1~kb files.
Note that for larger requests we could not correctly measure it
because client-side overheads affect its result a lot.

\noindent
\WC{2} \textbf{leveldb}.
We ran the default benchmark included in leveldb~\cite{leveldb}
(NoSQL-style DB), on the Mac mini.
\sys degrades the performance by 1-3\% and 0.3\% for
write and read, respectively.

\noindent
\WC{3} \textbf{Blogbench}.
We ran Blogbench~\cite{blogbench} on ext4 filesystem with the default configuration
in order to reproduce the load of a real-world busy file server.
\sys reported negligible overhead (0.2\%) for both write and read, on the Mac mini.

\begin{comment}
\PP{Image size}
We showed the code size increment
of \sys-enabled kernels (see~\autoref{table:binary-increase}):
Linux 5.12.0-rc1 for the Mac mini,
Linux 4.19.49 for the Raspberry Pi 3
and FreeBSD 11.0 for QEMU,
%
Overall, \sys increases about 5-9\%
of the image size,
e.g., increasing the number of instructions
and relevant metadata.
%
Note that the kernel image for the Raspberry Pi
contains instructions that emulate the performance of ARM PA.
\end{comment}

%
\begin{comment}
\autoref{table:instrumented}
shows how many of signing, authentication and resigning
happen in Linux and FreeBSD kernels.
% Re-signing for context changing is smaller than
% signing and authentication.
The difference between FreeBSD and Linux kernels is because
Linux's code, not whole binary, is about four times larger
than FreeBSD's.
\end{comment}

\begin{comment}
\begin{table}[t!]
\centering
\small
\begin{tabular}{lrr}
  \toprule
  PA operations & Linux 5.12.0-rc1  & FreeBSD \\
  \midrule
  Signing        & 1,830 & 2,612 \\
  %\quad Local    &  4,814 & 574 \\
  %\quad Global   & 33,468 & 2,038 \\
  Authentication & 2,860 & 1,296 \\
  Resigning      &  2,355 & 634 \\
  \bottomrule
\end{tabular}
\caption{The number of PA-related operations instrumented by \sys.}
\label{table:instrumented}
\end{table}
\end{comment}

\subsection{Context Analyzer}
\label{s:eval:analyzer}

\begin{table}[t!]
	\centering
	\scriptsize
	\begin{tabular}{@{}l ccccc@{}}
		\toprule
		\textbf{OS} & \multirow{1}{*}{\textbf{TAT}} & \multirow{1}{*}{\textbf{TDS at 10}} & \multirow{1}{*}{\textbf{TDS at 20}} & \multirow{1}{*}{\textbf{TDS at 30}} & \multirow{1}{*}{\textbf{TDS at 40}} \\
		\midrule
		\text{Linux} & 89 & 35/39.3\% & 50/56.1\% & 67/75.2\% & 80/89.8\% \\
		\text{FreeBSD} & 48 & 18/37.5\% & 30/62.5\% & 38/79.1\% & 41/85.4\% \\
		\bottomrule
	\end{tabular}
	\caption{The correlation result between allowed targets and diversity scores;
	TAT: the number of structs that rank top 10\% in allowed targets,
	TDS at $k$: the number of structs that rank top $k$\% in diversity scores;
	A higher TDS indicates the better.}
	\label{t:analyzer-sound}
\end{table}

We conducted in the following aspects an empirical evaluation for context analyzer.

\PP{Soundness}
The context analyzer is based on sound assumption that
the structures with the larger allowed targets
likely have the higher diversity scores.
To prove this assumption in practice,
we measured the correlation between allowed targets and diversity scores.
\autoref{t:analyzer-sound} shows that
the majority of TAT ranks top 20\% in diversity score,
which backs up our claim.

\PP{Security}
To see the security enhancement,
we measured how many contexts out of what rank top 10\% in allowed targets
could be successfully refined via \cc{objbind}/\cc{retbind}.
As a result, 272 out of 312 (87.1\%) and 350 out of 376 (93.0\%) could be resolved
for Linux and FreeBSD, respectively.

\PP{Failure cases}
We found several cases in which the context analyzer could not refine and why.
For both Linux and FreeBSD, we found the reason was mainly due to the absence of \cc{objtype},
which renders \cc{objbind} unapplicable.
Specifically, it falls into two cases--
1) local function pointers (e.g., \cc{fptr_t fp = func}) and
2) type casting (e.g., \cc{fptr_t fp = obj->fp}).
We plan to refine both cases by improving static analysis as future work.

\PP{Engineering efforts}
Despite the automation capability of the analyzer, minimal engineering efforts are still required
to deal with cases in which the diversity score is too low to apply (i.e., zero or one)
but the number of allowed targets is high.
Since the analyzer uses address binding for such cases, issues could arise from memory copy functions.
In the \autoref{table:allowed-target} setting, the task to deal with such issues took a day to complete by a person.
\section{Discussion}
\label{s:discussion}

\PP{Assurance over assembly code}
All assembly code is checked by the static validator---%
all inserted PA instructions
respect their security invariant (see~\autoref{ss:validator}).
However, it is still possible in theory
that \sys misses the protection of function pointers
generated and used in inline assembly.
Fortunately,
if such an assembly code ever just runs in \sys,
1) the system crashes immediately (\cc{aut} failure) in most cases,
so our exhaustive benchmarks help us address this issue,
and 2) if not crashed,
our run-time validator
helps us identify the problem
by scanning the entire memory for raw pointers.
We observed only one case
in \cc{kernel_thread} explained in~\autoref{s:runtime-validator}.

\begin{comment}
\PP{Automated approaches to insert annotation}
It would not be infeasible
to automate the discovery of \cc{objbind} targets
in theory.
However, we chose our design based on the principles of the Linux community,
which recommends explicit annotation
on security properties.
%
Any mistake in the automation process
introduces latent bugs that are difficult
to debug the root causes.
%
Our annotations,
similar to other developers' annotations (like \cc{__user}),
can be maintained together
with the code base
once in place.
%
% In the next study considering broader targets, we are planning
% to generalize and automate PAL’s approach.
\end{comment}

% \PP{Comparing with PA-based userspace protection}

\PP{Implication of reserving one PA key}
\sys reserved one PA key for the kernel protection
and another key for the user space.
However, this does not mean that
all user space applications
share the same key---%
each application has its own dedicated key
that is multiplexed by the kernel.
%
\begin{comment}
One might consider using PA
for \emph{two} protection domains
in one application,
but we haven't observed
any such usage in user space applications or
their security benefits
over a single domain.
\end{comment}

\begin{comment}
\PP{Combined instructions for atomicity}
To make it atomic that uses of PA illustrated in \autoref{fig:nonatomic},
ARM provides combined instructions named \cc{blraa/blrab} that combine authentication and branch,
for the first case in \autoref{fig:nonatomic} but nothing for the second and third case.
As both cases are frequently used,
our static validator (\autoref{ss:validator}) has to validate the final image
unless ARM introduces new combined instructions for all of uses of PA.
\end{comment}

\PP{Denial-of-Service}
In terms of security,
\sys should panic
at any authentication failure.
However, in respect of
the recommended policy on security violations in Linux
community~\cite{linus-security},
\sys provides a better alternative
that any vendor can enable based on their goal.

% Mitigating brute-force attacks through delaying (see~\autoref{s:brute-force-attack})
% does not mean it is possible to eradicate DoS.
% But, unlike crashing, delaying can prevent from DoS on the cases
% that are not malicious attempts but just unexpected failures (because such events happen rarely).
% This design is based on

\begin{comment}
\PP{Maximizing CFI precision}
Manual annotation (i.e., forcely binding to the function pointer's address)
can enhance the precision
even in the place having too low diversity score (i.e., zero or one)
to generate annotation by the context analyzer.
%
This approach would cause some side-effects
(i.e., address changes by \cc{memcpy}),
but fortunately, resolvable for skillful developers
with adequate test tools.
%
Accordingly, \sys takes account of usability to trade off diversity scores
against automation capability.

\PP{False positive of static validator}
Since it is infeasible to implement perfect binary analysis on the kernel,
our static validator in initial development stage
reported about 100 false positives but no false negatives.
They arose mainly due to complicated control flows in the kernel and are confirmed by hand.
This task took about two days by a person.
Since this is burdensom for developers, we plan to improve our static validator
to reduce the number of false positives.
\end{comment}

% \PP{Remaining attack vectors}
% Once CFI protections are deployed,
% an adversary would like to consider
% data-oriented attacks~\cite{dfi}.
% %
% \XXX{think more about.. tomorrow.i}

\section{Related Work}
\label{s:relwk}

Ever since a CFI-based approach was introduced to mitigate
code-reuse attacks~\cite{cfi},
a number of research ideas
have been proposed to improve
its protection precision
and run-time performance~\cite{cfi-survey}.
%
% Notably, online approaches~\cite{picfi, ding:pittypat}
% to estimating the indirect-call targets during execution
% effectively refine the target set
% to \emph{one}, so-called $\mu$CFI~\cite{hu:ucfi}.
%
Since precision and performance
are fundamental trade-offs in CFI,
the finest target estimation
comes with non-negligible performance overhead,
rendering them unattractive for practical adoption.
In contrast,
the coarse-grained CFI solutions,
like Microsoft's Control-flow Guards~\cite{ms-cfg},
Google's Indirect Function-Call Checks~\cite{android-cfi},
PaX's Reuse Attack Protector (RAP)~\cite{pax-rap},
and Apple's PA~\cite{apple-cfi},
have been successfully deployed
to protect web browsers and operating systems.

% In this section, we relate
% our approach to existing research
% in two dimensions, namely,
% the hardware primitives
% in accelerating the CFI-relevant operations,
% and comparing ours
% with other in-kernel CFI approaches.

\PP{Hardware-based CFI}
Silicon-level features
can significantly alleviate
the performance overhead of CFI.
For example,
commodity technologies
have been used to
design lightweight CFI schemes:
Intel PT~\cite{intel-pt}
to trace control-flow changes~\cite{ding:pittypat,ucfi},
Intel LBR~\cite{lwn-lbr}
to get the history of branch changes~\cite{kbouncer,roppecker},
% and similarly for PMU~\cite{intel-pmu}
% to monitor system events at run-time~\cite{cfimon};
and Intel MPX~\cite{intel-mpx}
to quickly enforce target boundaries~\cite{ocfi}.
Since these hardware features are not intended
for security,
retrofitting them for CFI
leaves a lot of weaknesses in security
like PT packet losses~\cite{ding:pittypat,ucfi}
or overflowing branch history~\cite{kbouncer}.

Recently,
more hardware primitives%
~\cite{pointer-tagging, custom1, custom2, custom3, custom4, intel-cet, arm-8-6}
are designed specifically to assist CFI---%
we use the term, ``primitives,''
as they are dependent
on the software counterpart
that utilizes the primitive for the full protection.
%
% For example, Intel CET
% provides a fast backward-edge protection~\cite{intel-cet},
% but requires a software-based forward-edge solution
% for full CFI protection.
%
% Similarly,
% ARM's BTI~\cite{arm-8-6, arm-reference}
% provides machinery to explicitly annotate
% the branch targets
% as part of the compilation process.
%
% Although the aforementioned technologies look promising,
% both are not yet available in commodity chips.
%
ARM's PA~\cite{qualcomm-pac}
is one of the most promising primitives
that Apple first utilized
to enforce CFI in iOS and M1-based macOS%
~\cite{apple-cfi}.
In academia,
PARTS~\cite{pacitup} and PATTER~\cite{patter}
also proposed type-based signing by using PA,
but hardly beyond the intended design of PA~\cite{pacitup}.
Apple's CFI implemented
much advanced type analysis
to address unique challenges to its own kernel---%
mixed uses of objective-C and native components~\cite{apple-cfi,ios-mac-security}.
Unfortunately,
Apple's approach is not universally applicable
to other monolithic commodity OSes
like Linux and FreeBSD
in providing finer-grained target enforcement
for CFI (\autoref{s:eval:security}).

\PP{In-kernel CFI}
% The unique structure of system software
% like the kernel or hypervisor
% exhibits non-trivial challenges
% to enforce valid control-flows
% (e.g., context switching).
% %
Commercial solutions
such as
PA-based CFI for iOS~\cite{apple-cfi},
LLVM's CFI for Android~\cite{androidblog-cfi},
and PaX RAP for Linux~\cite{pax-rap}
use the type-based approach
to refine the precision of CFI
without breaking code-level compatibility.
Academic approaches
have explored
various directions
to further enhance its precision,
by utilizing mapping tables derived from finer-grained CFGs,
for system software~\cite{kcfi, hypersafe, kcofi}.
Our approach,
while providing a commercial solution,
aims to achieve
the finest precision
with minimal performance overheads
on commodity hardware supporting PA.

\section{Conclusion}
\label{s:conclusion}
This paper presents \sys,
an in-kernel, ARM PA-based protection
that enhances the precision of CFI
with minimal performance overhead.
We define new attack vectors
for PA when used to protect the kernel
and found erroneous cases
in the state-of-the-art PA-based protections
such as iOS and PARTS.
\sys provides two techniques:
automated refinement techniques
to capture idioms and design patterns for better CFI,
and a static validator to check error-prone usage patterns
of PA in the final OS images.
\sys has been ported to Linux and FreeBSD
and our evaluation shows
negligible performance overhead.
We will make \sys publicly available
upon acceptance and for artifact evaluation.

% \input{ack}

%\balance
\bibliographystyle{plain}
\footnotesize
\bibliography{p,sslab,conf}

\clearpage
\appendix

\pdfbookmark{\appendixname}{Appendix}
\section*{Appendix}

\hypersetup{bookmarksdepth=-1}

\section{Context Analyzer Algorithm}
\label{appendix:analyzer-algo}

\autoref{a:analyzer-algo} introduces \textsc{estimate_diversity_score},
a core algorithm to estimate diversity score (shortly, DS)
for the context analyzer annotating \cc{objbind}.

The algorithm takes inputs --
$funcs$ (all functions in kernel), $struct$ and $fidx$ (a desired target structure and its field index for DS),
and breaks down to three phases :

1) It collects all assignments (i.e., store instructions) to the given $struct.fidx$ (line 21)
and performs the Andersen's pointer analysis to the value operand of each assignment (line 22)
, which retrieves points-to-set ($pts$) of the value operand (line 23).
The analysis is a flow-intensive and path-insensitive intra-procedural analysis.

2) It attempts to resolve all $pts$ retrieved from the previous phase by running $check(pts)$ function
that checks if all value in $pts$ meets any of the conditions to increment diversity score (see, \autoref{ss:context-analyzer})
and returns true if that is the case. (line 25)
If at least one $pts$ fails in $check(pts)$, it moves off to the third phase where an iterative inter-procedural analysis plays.

3) If the else branch at line 28 is taken, it starts to run an iterative worklist algorithm (line 29 ~ line 44).
This algorithm finds and adds functions that need to be further investigated (i.e., calling contexts) into $wk$ in an iterative way,
unless there is nothing in $wk$ or the number of rounds goes over the threshold we set (five)
to avoid being unterminated caused by the large code size of the kernel.
To find such functions, it checks if a value in $pts$ is used as argument in a function call to $f$ in $wk$ (line 37),
and if true it attempts to increment diversity score if possible (line 38)
otherwise adds the current context into $upd$ (line 41) to repeat this algorithm.

After all of the above phases are complete, it finally returns DS on $struct.fidx$ (line 45).

\begin{figure}
\footnotesize
\SetKwFunction{validate}{\textsc{validate_bb}}
\SetKwInOut{Symbol}{Symbol}
\SetKwInOut{Input}{Input}
\SetKwInOut{Output}{Output}
\SetKwProg{EstimateDiversityScore}{\textsc{estimate_diversity_score}}{}{}
\nonl\EstimateDiversityScore{($funcs,struct,fidx$)}{
  \Input{$funcs$: all functions in kernel\newline
  $struct$: a target structure\newline
  $fidx$: a target field index}
  \Output{diversity score}
  \Symbol{
  $f.insns$: all instructions in $f$\;\newline
  $si$: store instruction\;\newline
  $ds$: diversity score\;\newline
  $pts$: points-to set\;\newline
  $wk$: current worklist (map<$f$,$pts$>)\;\newline
  $upd$: worklist for next round\;\newline
  $first\_wk$: worklist for first round\;\newline
  $apts(f,p)$: andersen pointer analysis on $p$ within $f$\;\newline
  $check(pts)$: check if $pts$ meets $ds$ conditions\;}
  \vspace{2px}
	$ds \gets 0$\;

	$wk \gets \varnothing$\;
	$first\_wk \gets \varnothing$\;

  \ForEach{$f \in funcs;$}{
	// 1. set up the first worklist

	\ForEach{$i \in f.insns;$}{
		\If{$i = si \And i.pointer\_op = struct.fidx$}{
			$pts \gets apts(f, i.value\_op)$\;

			$first\_wk.add(f, pts)$\;
		}
	}
	\ForEach{$f,pts \in first\_wk;$}{
		\If{$check(pts) = true$}{
			// 2. increment ds if possible

			$ds \gets ds + 1$\;
		}
		\Else{
			// 3. start iterative worklist algorithm

			$depth \gets 0$\;

			$wk \gets first\_wk$\;

			\While{$wk.size() > 0 \And depth < 5$} {
				\ForEach{$f,pts \in wk;$}{
					// iterate instructions in funcs

					\ForEach{$i \in funcs.insns;$}{
						\If{$i$ is a call to $f \And i.arg \in pts$} {
							$pts \gets apts(f, i.arg)$\;

							\If{$check(pts) = true$}{
								$ds \gets ds + 1$\;
							}
							\Else {
								$upd.add(i.func, i.arg)$\;
							}
						}
					}
				}

				$wk \gets upd$\;

				$upd \gets \varnothing$\;

				$depth \gets depth + 1$\;
			}
		}
	}
  }
  \Return $ds$\;
}
\caption{
	The core algorithm to estimate a diversity score for \cc{objbind}.
}
\label{a:analyzer-algo}
\end{figure}

\clearpage

\section{Abusing Preemption Context as Signing Oracle}
\label{appendix:preempt-attack}

\begin{figure}[!h]
  \includegraphics[page=1, width=1.1\columnwidth]{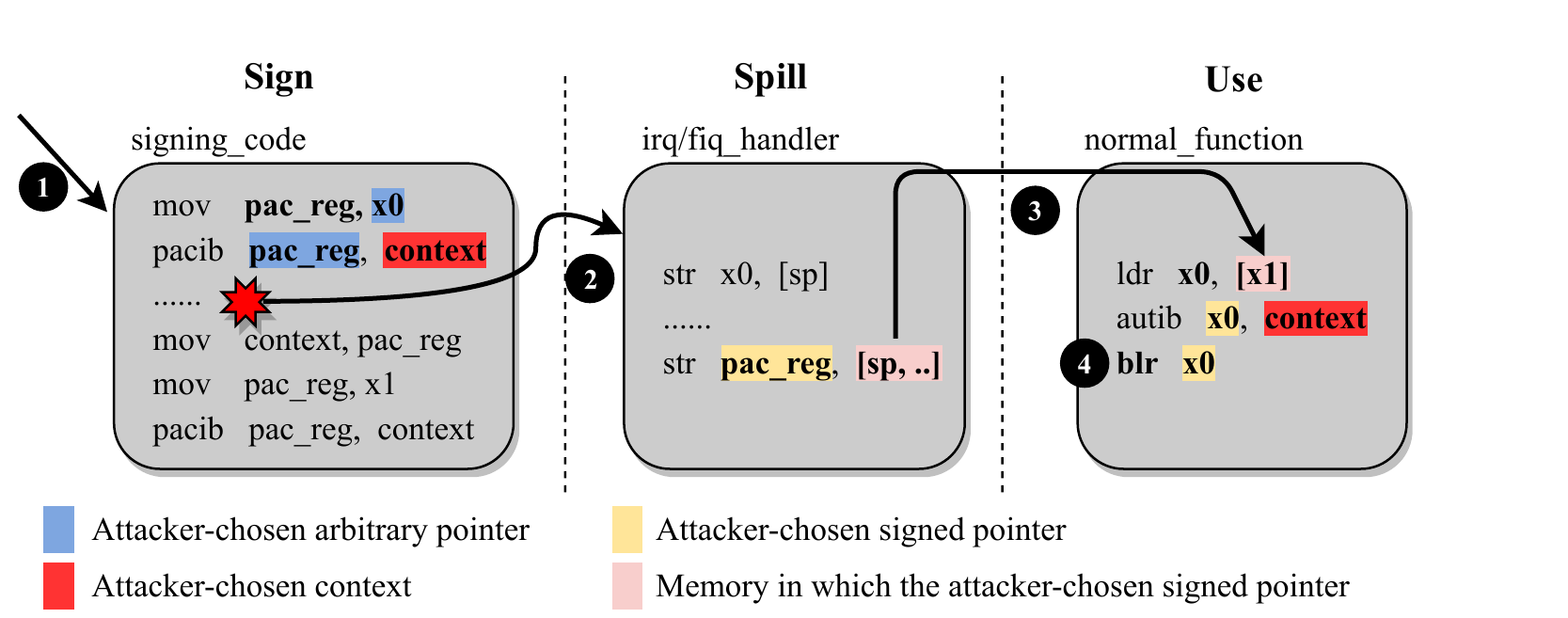}
  \caption{
  Exploiting the code for signing the preemption context as a signing oracle.
  \protect\BC{1} enter the signing routine via IRQ or creating thread state (e.g., \cc{arm_saved_state_t} in iOS) and
  sign an attacker-chosen pointer in the first register \cc{x0} with an attacker-chosen context.
  \protect\BC{2} preempt the signing via IRQ/FIQ and spill the signed pointer onto the stack memory
  (FIQ is the high-priority interrupt, which can preempt IRQ in arm64).
  \protect\BC{3} read the pointer from the spilled stack memory and substitute the pointer for an indirect call that uses the attacker-chosen context.
  \protect\BC{4} Consequently an attacker is able to jump to the attacker-chosen place. (i.e, \cc{x0} in \protect\BC{1})}
  \label{fig:preemptattack}
\end{figure}

\clearpage

\section{Evaluation Supplement Data}
\label{appendix:evaluation}

\setcounter{table}{0}
\renewcommand{\thetable}{C\arabic{table}}

\begin{table} [h!]
	\centering
	\tiny
	\resizebox{\textwidth}{!}
	{\begin{tabular}{lrrrrrrrr}
		\toprule
		&
		& \multicolumn{3}{c}{\textbf{4.19.49 on Rpi3}}
		&
		& \multicolumn{3}{c}{\textbf{5.12.0-rc1 on Mac mini(M1)}} \\
		\midrule
		&
		& \multicolumn{1}{c}{\textbf{Stock (sec)}}
		& \multicolumn{1}{c}{\textbf{w/ \sys (sec)}}
		& \multicolumn{1}{c}{\textbf{Overhead}}
		&
		& \multicolumn{1}{c}{\textbf{Stock (sec)}}
		& \multicolumn{1}{c}{\textbf{w/ \sys (sec)}}
		& \multicolumn{1}{c}{\textbf{Overhead}} \\
		\midrule
		{SPEC} & 400.perlbench & 1.631 & 1.634 & 0.003 /\phantom{+}0.18\% & & \multicolumn{3}{c}{-} \\
		{ 2006}& 401.bzip2 & 4.616 & 4.615 & -0.001 / -0.02\% & & \multicolumn{3}{c}{-} \\
		& 403.gcc & 12.701 & 12.68 & -0.021 / -0.17\% & & \multicolumn{3}{c}{-} \\
		& 429.mcf & 19.082 & 19.085 & 0.003 /\phantom{+}0.02\% & & \multicolumn{3}{c}{-} \\
		& 435.gromacs & 9.769 & 9.775 & 0.006 /\phantom{+}0.06\% & & \multicolumn{3}{c}{-} \\
		& 436.cactusADM & 30.885 & 31.07 & 0.185 /\phantom{+}0.60\% & & \multicolumn{3}{c}{-} \\
		& 444.namd & 118.309 & 118.394 & 0.085 /\phantom{+}0.07\% & & \multicolumn{3}{c}{-} \\
		& 445.gobmk & 1.329 & 1.33 & 0.001 /\phantom{+}0.08\% & & \multicolumn{3}{c}{-} \\
		& 447.dealII & 168.277 & 168.773 & 0.496 /\phantom{+}0.29\% & & \multicolumn{3}{c}{-} \\
		& 456.hmmer & 3.275 & 3.268 & -0.007 / -0.21\% & & \multicolumn{3}{c}{-} \\
		& 458.sjeng & 23.691 & 23.655 & -0.036 / -0.15\% & & \multicolumn{3}{c}{-} \\
		& 462.libquantum & 0.324 & 0.324 & 0.000 /\phantom{+}0.00\% & & \multicolumn{3}{c}{-} \\
		& 464.h264ref & 151.026 & 150.628 & -0.398 / -0.26\% & & \multicolumn{3}{c}{-} \\
		& 470.lbm & 42.153 & 42.143 & -0.010 / -0.02\% & & \multicolumn{3}{c}{-} \\
		& 471.omnetpp & 3.751 & 3.75 & -0.001 / -0.03\% & & \multicolumn{3}{c}{-} \\
		& 473.astar & 63.975 & 63.839 & -0.139 / -0.20\% & & \multicolumn{3}{c}{-} \\
		& 483.xalancbmk & 0.937 & 0.936 & -0.001 / -0.11\% & & \multicolumn{3}{c}{-} \\
		& 999.specrand & 0.113 & 0.113 & 0.000 /\phantom{+}0.00\% & & \multicolumn{3}{c}{-} \\
		\bottomrule
	\end{tabular}}
	\parbox{\textwidth}{
		\vspace{5pt}
		\caption{Detailed data about overhead of SPEC2006.}
	}
	\label{t:benchmark-spec}
\end{table}

\begin{table} [h!]
	\centering
	\tiny
	\resizebox{\textwidth}{!}
	{\begin{tabular}{lrrrrrrrr}
		\toprule
		&
		& \multicolumn{3}{c}{\textbf{4.19.49 on Rpi3}}
		&
		& \multicolumn{3}{c}{\textbf{5.12.0-rc1 on Mac mini(M1)}} \\
		\midrule
		&
		& \multicolumn{1}{c}{\textbf{Stock ($\mu$s)}}
		& \multicolumn{1}{c}{\textbf{w/ \sys ($\mu$s)}}
		& \multicolumn{1}{c}{\textbf{Overhead}}
		&
		& \multicolumn{1}{c}{\textbf{Stock ($\mu$s)}}
		& \multicolumn{1}{c}{\textbf{w/ \sys ($\mu$s)}}
		& \multicolumn{1}{c}{\textbf{Overhead}} \\
		\midrule
		{LMbench} & null & 2.38 & 2.64 & 0.26 /10.9\% & & 0.1489 & 0.1971 & 0.0482 /\phantom{+}32.4\%\\
		& fstat & 3.70 & 3.97 & 0.27 /\phantom{+}7.3\% & & 0.8170 & 0.2837 & -0.5342 / -65.3\%\\
		& open_close & 31.46 & 34.24 & 2.78 /\phantom{+}8.8\% & & 1.0315 & 1.1890 & 0.1575 /\phantom{+}15.2\%\\
		& select_200 & 33.65 & 31.70 & -1.95 / -5.8\% & & 19.0521 & 2.5541 & -16.4980 / -86.6\%\\
		& sig_install & 4.92 & 5.44 & 0.52 /10.6\% & & 0.7392 & 0.1989 & 0.2460 /\phantom{+}23.7\%\\
		& sig_catch & 29.36 & 31.70 & 2.34 /\phantom{+}8.0\% & & 7.2327 & 1.0529 & -6.1798 / -85.4\%\\
		& protection_fault & 0.30 & 0.60 & 0.30 / 100\% & & 0.8306 & 0.2222 & -0.6084 / -73.2\%\\
		& pipe & 69.05 & 73.47 & 4.42 /\phantom{+}6.4\% & & 17.3443 & 19.3347 & 3.2874 /\phantom{+}11.5\%\\
		& unix_sock & 84.28 & 89.27 & 4.99 /\phantom{+}5.9\% & & 18.0461 & 18.2759 & 0.2298 /\phantom{+}1.27\%\\
		& fork_exit & 719.78 & 746.14 & 26.36 /\phantom{+}3.7\% & & 85.6061 & 91.7627 & 6.1566 /\phantom{+}7.19\%\\
		& fork_exec & 774.40 & 802.43 & 28.03 /\phantom{+}3.6\% & & 101.3725 & 103.0943 & 1.7218 /\phantom{+}1.70\%\\
		\midrule
		{Linux}& \cc{messaging} & 2.977 sec & 3.089 sec & 0.112 / 3.76\% & & 0.164 sec & 0.169 sec & 0.005 / 3.0\%\\
		{perf}& \cc{pipe} & 69.603 sec & 73.212 sec & 3.609 / 5.19\% & & 18.087 sec & 18.941 sec & 0.854 /\phantom{+}4.7\%\\
		\bottomrule
	\end{tabular}}
	\parbox{\textwidth}{
		\vspace{5pt}
		\caption{Detailed data about overhead of Lmbench, perf-sched benchmarks.}
	}
	\label{t:benchmark-micro}
\end{table}

\clearpage
\begin{table} [h!]
	\centering
	\tiny
	\resizebox{\textwidth}{!}{
	\begin{tabular}{lrrrrrrrr}
		\toprule
		&
		& \multicolumn{3}{c}{\textbf{4.19.49 on Rpi3}}
		&
		& \multicolumn{3}{c}{\textbf{5.12.0-rc1 on Mac mini(M1)}} \\
		\midrule
		&
		& \multicolumn{1}{c}{\textbf{Stock (ms)}}
		& \multicolumn{1}{c}{\textbf{w/ \sys (ms)}}
		& \multicolumn{1}{c}{\textbf{Overhead}}
		&
		& \multicolumn{1}{c}{\textbf{Stock (ms)}}
		& \multicolumn{1}{c}{\textbf{w/ \sys (ms)}}
		& \multicolumn{1}{c}{\textbf{Overhead}} \\
		\midrule
		{apache} & 1~KB & 3.13 & 3.16 & 0.03 / 1.06\% & & 0.132 & 0.133 & 0.001 / 0.75\%\\
		& 10~KB & 4.10 & 4.12 & 0.02 / 0.46\% & & \multicolumn{3}{c}{-} \\
		& 100~KB & 12.00 & 12.00 & 0.00 / 0.02\% & & \multicolumn{3}{c}{-} \\
		& 200~KB & \multicolumn{3}{c}{-} & & \multicolumn{3}{c}{-} \\
		& 1~MB & 92.57 & 92.64 &  0.08 / 0.08\% & & \multicolumn{3}{c}{-} \\
		& 10~MB & 895.87 & 895.78 &  0.10 / 0.01\% & & \multicolumn{3}{c}{-} \\
		\midrule
		{leveldb} & fillseq & \multicolumn{3}{c}{-} & & 1.692 & 1.745 & 0.053 /\phantom{+}3.10\%\\
		& fillsync & \multicolumn{3}{c}{-} & & 6.861 & 6.990 & 0.129 /\phantom{+}1.80\%\\
		& fillrandom & \multicolumn{3}{c}{-} & & 4.892 & 5.013 & 0.121 /\phantom{+}2.40\%\\
		& overwrite & \multicolumn{3}{c}{-} & & 4.869 & 4.665 & -0.204 / -4.10\%\\
		& readrandom & \multicolumn{3}{c}{-} & & 9.332 & 9.368 &  0.036 /\phantom{+}0.38\%\\
		& readseq & \multicolumn{3}{c}{-} & & 0.573 & 0.575 & 0.002 /\phantom{+}0.34\%\\
		& readreverse & \multicolumn{3}{c}{-} & & 1.075 & 1.069 & -0.006 / -0.55\%\\
		\midrule
		{blogbench} & write & \multicolumn{3}{c}{-} & & 382 & 381 & 1 / 0.2\%\\
		& read & \multicolumn{3}{c}{-}  & & 416368 & 415550 & 818 / 0.2\%\\
		\bottomrule
	\end{tabular}}
	\parbox{\textwidth}{
		\vspace{5pt}
		\caption{Detailed data about overhead of Apache, Leveldb, Blogbench benchmarks. The blogbench's results are based on throughput.}
	}
	\label{t:benchmark-macro}
\end{table}

\begin{table}[h]
	\footnotesize
	\centering
	\begin{tabular}{lcccc}
		\toprule
		& 5.12.0-rc1/the Mac mini & 4.19.49/Rpi3 & FreeBSD/Qemu \\
		\midrule
		Stock & 123.5 MB & 19.9 MB & 5.9 MB\\
		w/ \sys & 130.7 MB & 23.0 MB  & 6.4 MB \\
		Overhead & 7.2 / 5.8\% & 3.1 / 15.6\% & 0.5 / 8.5\%\\
		\bottomrule
	\end{tabular}
	\caption{Image sizes increased by \sys in Linux and FreeBSD kernels.}
	\label{table:binary-increase}
\end{table}

\end{document}